\documentclass[10pt,journal,compsoc]{IEEEtran}
\ifCLASSOPTIONcompsoc
  \usepackage[nocompress]{cite}
\else
  \usepackage{cite}
\fi
\usepackage{amsmath,amssymb,amsfonts}
\usepackage{pifont}
\usepackage{bbding}
\usepackage{pifont}
\usepackage{algorithm}
\usepackage{algpseudocode}
\usepackage{graphicx}
\usepackage{tabularx}
\usepackage{caption}
\usepackage{subcaption}
\usepackage{textcomp}
\usepackage{xcolor}
\newcommand{\xmark}{\ding{55}}
\newcommand{\rakib}[1]{\textcolor{black}{#1}}
\newcommand{\rakibtc}[1]{\textcolor{black}{#1}}
\newcommand{\tcad}[1]{\textcolor{black}{#1}}

\newcommand{\tcrevision}[1]{\textcolor{black}{#1}}
\ifCLASSINFOpdf
\else
\fi
\begin{document}
%
\title{ A Neural Network-based SAT-Resilient Obfuscation Towards Enhanced Logic Locking}

\author{Rakibul Hassan,~\IEEEmembership{Student Member,~IEEE}, 
Gaurav Kolhe,~\IEEEmembership{Student Member,~IEEE},
Setareh Rafatirad, ~\IEEEmembership{Senior Member,~IEEE}, 
Houman Homayoun,~\IEEEmembership{Senior Member,~IEEE} and \\   Sai Manoj Pudukotai Dinakarrao,~\IEEEmembership{Member,~IEEE} \\
\thanks{R. Hassan, and S. M. P. Dinakarrao are associated with Department of Electrical and Computer Engineering, George Mason University, Fairfax, VA 22030, USA. Email: \{rhassa2,spudukot\}@gmu.edu}
\thanks{G. Kolhe, and H. Homayoun are associated with Department of Electrical and Computer Engineering, University of California, Davis, Davis, CA 92617, USA. Email: \{gskolhe,hhomayoun\}@gmu.edu}
\thanks{S. Rafatirad is associated with Department of Information Sciences and Technology, George Mason University, Fairfax, VA 22030, USA. Email: srafatir@gmu.edu}
\thanks{A preliminary version of this work is accepted in ISQED 2020 for publication. 
}}

\IEEEtitleabstractindextext{%
\begin{abstract}
Logic obfuscation is introduced as a pivotal defense against multiple hardware threats on 
Integrated Circuits (ICs) including reverse engineering (RE) and intellectual property (IP) theft. The effectiveness of logic obfuscation is challenged by 
recently introduced Boolean satisfiability (SAT) attack and 
it's variants. 
A plethora of counter measures have also been proposed to thwart the SAT attack. Irrespective of the implemented defense against SAT attacks, large power, performance and area overheads are seen to be indispensable. 
In contrast, 
we propose a cognitive solution which is a neural network based unSAT clause translator, SATConda, that incurs a minimal area and power overhead while preserving the original functionality with impenetrable security. 
SATConda is incubated with a unSAT clause generator that 
translates the existing conjunctive normal form (CNF) through minimal perturbations such as 
inclusion of pair of inverters or buffers or adding new 
lightweight unSAT block depending on the provided CNF. 
For efficient unSAT clause generation, SATConda is equipped with 
a multi-layer neural network that first learns the dependencies of features (literals and clauses), followed by 
a long-short-term-memory (LSTM) network to validate and backpropagate the SAT-hardness for better learning and translation. 
Our proposed SATConda is evaluated on ISCAS'85 and ISCAS'89 benchmarks and is seen to successfully defend against multiple state-of-the-art SAT attacks devised for hardware RE. 
In addition, we also evaluate our proposed SATConda's empirical performance against MiniSAT, Lingeling and Glucose SAT solvers 
that form the base for numerous existing deobfuscation SAT attacks.

\end{abstract}

}

\maketitle

\IEEEdisplaynontitleabstractindextext

%
\IEEEpeerreviewmaketitle

\IEEEraisesectionheading{\section{Introduction}\label{sec:introduction}}

%
%
%
%
With the semiconductor industries inclining 
towards fabless business model i.e., outsourcing the fabrication to offshore foundries to cope-up with 
the operational and maintenance costs, 
the hardware security threats are exacerbating. This hardware threat could be in any form including 
intellectual property (IP) theft, 
integrated circuit (IC) tampering, over production and cloning \cite{Karri_CS'10,Rostami_Proc_IEEE_14}. 
What is worse, the threat could occur during any 
phase of the IC production cycle ranging from 
design phase, fabrication phase or even after 
releasing the design to the market 
(in form of side-channel attacks) 
.

To thwart the prevalent security threats, many hardware design-for-trust techniques 
have been introduced such as split manufacturing \cite{Rajendran_DATE'13}, IC 
camouflaging, and logic locking \emph{a.k.a} 
logic obfuscation \cite{Yasin_TCAS'15}. Among 
multiple aforementioned techniques, logic locking can thwart the majority of the attacks 
at various phases in the IC Production chain \cite{Yasin_HOST'16}. This is because
logic locking 
requires the correct keys to unlock the true functionality of the design. Additionally, 
as a part of the post-manufacturing process, the activation of IC 
(i.e., providing correct keys) will be accomplished in a trusted regime to hide 
the functionality from the untrusted foundry and other attacks. Having 
key-programmable gates allows the designer or user to control the functionality 
using these key inputs.

Although logic locking schemes enhance the security of the IP, 
the advent of Boolean satisfiability (SAT) based attack \cite{Subramanyan_HOST'15}, 
also known as ``oracle-guided'' threat model shows that by applying 
stimuli to the design and analyzing the output, the key value and functionality 
of an IC could be extracted in the order of a few minutes or less. To implement SAT attack, 
the attacker needs access to (a) an obfuscated netlist of IC 
(obtained after 
de-layering IC or constructed from layout), and (b) a 
functional/activated IC, to which the attacker can apply stimuli and monitor the output. 
The extracted netlist is converted into a conjunctive normal form (CNF\footnote{A CNF  is a conjunction (i.e., AND) of one or more clauses, where a clause is a disjunction (i.e., OR) of literals.}), fed to a SAT solver to determine the keys (assignment to each Boolean variable or literal in the CNF) to decrypt and reverse engineer (RE) the IC/IP. 
It has been seen that modern SAT solvers can solve a SAT-problem 
with up to million variables \cite{franco2009handbook}. 

To mitigate SAT attack  several logic locking \cite{Yasin_HOST'16} techniques have been proposed. 
A recently proposed mechanism on logic locking was presented to mitigate SAT attack by introducing an additional logic block that makes SAT attack computationally infeasible \cite{Xie_CHES'16}. Recent literature reported signal probability skew (SPS) attack \cite{Yasin_ASP-DAC_17} against Anti-SAT defense \cite{Xie_CHES'16} which can break the Anti-SAT defense within few minutes.

One of the major challenges in adopting the existing defenses against SAT attacks or its variants 
is the imposed overheads in terms of area and power with no guarantee of security 
\cite{Kolhe_GLSVLSI'19}. 
Previous works \cite{Xie_CHES'16,Yasin_ASP-DAC_17} consider developing Anti-SAT solutions 
through embedding different metrics (properties of netlist that cannot be translated into CNF) 
or through heuristic intuitions. 
Such defenses involve 
challenges including complexity, incompleteness and high probability to exclude parameters that were not explored in literature. 
To address these concerns, we introduce SATConda\footnote{SATConda is our proposed defense mechanism to fight against the SAT-attack. The source code is available at https://github.com/hidden-for-anonymity }, 
equipped with a CNF 
generator that can 
convert the provided SAT prone
CNF into unSAT through minimal modification to the netlist such as flipping a literal by adding one or few gates 
(through addition of inverter gate or using XNOR instead of XOR are some of the na\"ive possibilities) 
in a clause of CNF i. e., converts the SAT distribution to unSAT distribution by learning the distributions yet preserving the functionality. 
To perform such modifications, we deploy neural network with 
bipartite message passing mechanism to 
cognitively learn and determine the properties of a CNF and distinguish 
SAT and unSAT problems. 
Once learnt, the amount of clauses added or the perturbations are introduced cognitively, which 
can be controlled to determine the trade-off between overheads and security. The two main contributions of this work are:


\begin{itemize}
   \item 
    SATConda induces additional clauses or flips the existing clauses to make the CNF (obfuscated IC) SAT hard. To perform such an operation cognitively, SATConda utilizes a neural network model to learn the distinguishing features of SAT and unSAT CNFs. 
    SATConda seeds learned parameter to the clause generator and then integrates that unSAT block to the original circuit in a way that the SAT attack fails to decrypt the keys used for the encryption. 
    \item 
    We successfully defend the standard benchmarks against existing attacks such as SAT-attack\cite{Subramanyan_HOST'15} by introducing an unSAT block and encrypting that block the the original circuit. 
    Using SATConda, we showcase the existing obfuscation schemes can be made robust with minimal modifications. 
    \end{itemize}
To the best of our knowledge the proposed technique is a novel defense mechanism against SAT-attack utilized 
neural network model to learn the SAT and unSAT clauses and encrypt the circuit under minimal overheads constraint. The idea of exploring deep neural network in circuit obfuscation, especially for converting SAT to unSAT is unexplored and is novel \cite{hassan2020satconda}. 
We also evaluate the translated obfuscated circuit with three different state-of-the-art SAT solvers. 
In addition to evaluating the SAT hardness, we have evaluated the area and power overheads incurred with additional security deployment through SATConda. 
\vspace{-0.5em}

\section{Background} 

Here, we discuss the basic information regarding the logic locking and the SAT attack. 
\subsection{Logic Locking}
Logic locking mechanism is implemented in a design by adding additional gates a.k.a ``key-gates'' to secure the circuit (IC/IP) 
by inducing the randomness in the observable output \cite{Yasin_TCAS'15}. To achieve the desired output from the design, all the key-gates must be set to their proper input. Any incorrect insertion to any of the key-gates leads to the incorrect output. Thus, an attacker needs to know the correct assignment 
to those keys-gates to decode the actual functionality of the design.

\begin{figure}
  \centering
  \begin{subfigure}[t]{.48\linewidth}
    \includegraphics[width=\linewidth]{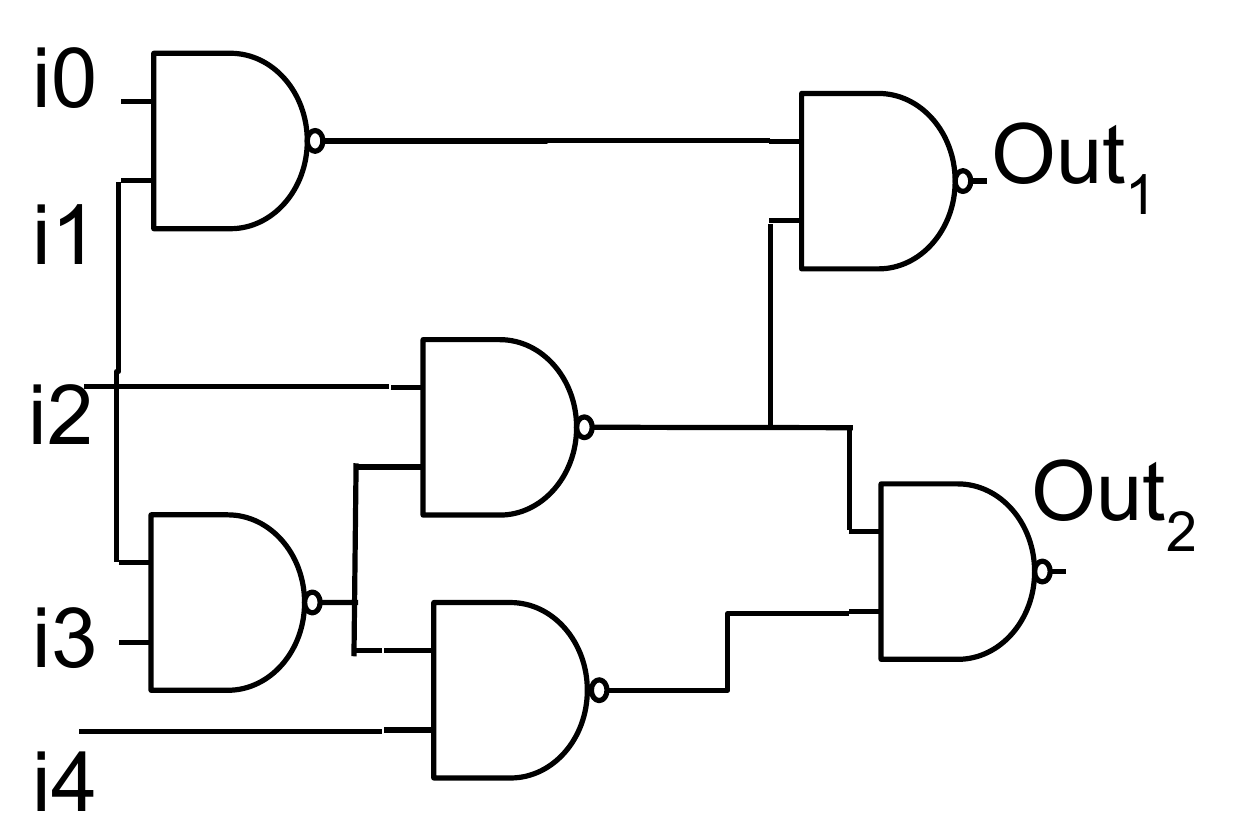}
    \caption{}
    \label{fig:c17-ori}
  \end{subfigure}
  \begin{subfigure}[t]{.5\linewidth}
    \includegraphics[width=\linewidth]{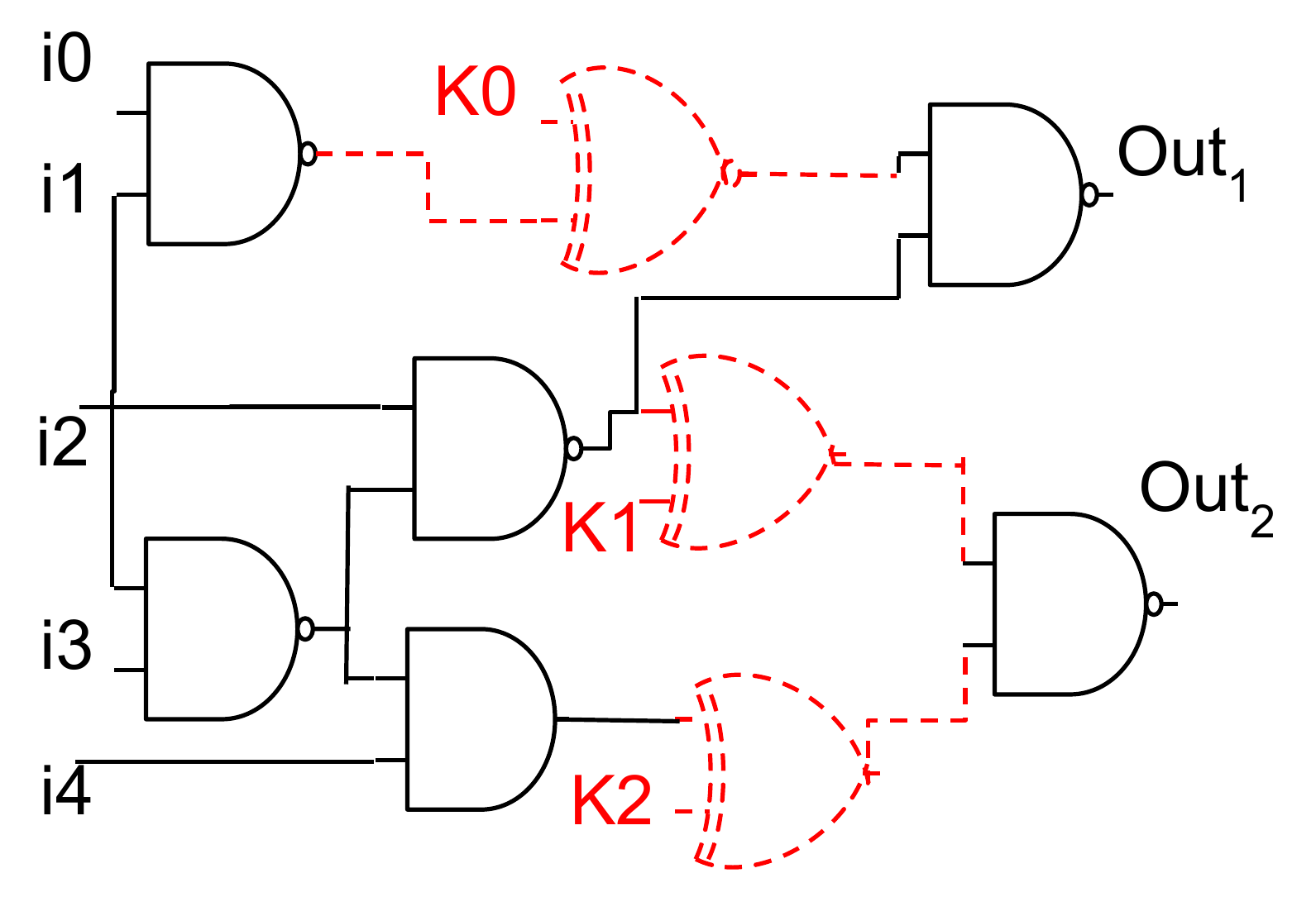}
    \caption{}
    \label{fig:c17-encrypted}
  \end{subfigure}
  \caption{Logic locking on c17 benchmark circuit. (a) Original Circuit Design, (b) Encrypted circuit with additional key-gates (dotted lines/gates) \cite{Dupuis_JET'19}.  Desired circuit behavior is achieved when (k2, k1, k0) = 101 }
  \label{fig:c17-logic-lock}
\end{figure}

Figure \ref{fig:c17-ori} depicts the original circuit and the 
Figure \ref{fig:c17-encrypted} shows a logic locked circuit of the same 
C17 circuit from ISCAS'85 benchmark \cite{Hansen_DES'99}. The original circuit consists of five inputs with six NAND gates and two outputs. 
The encryption is done by adding three additional gates, termed 
as key-gates. 
For Figure \ref{fig:c17-encrypted}, if one assigns (k2, k1, K0) as (1, 0, 1), only then the circuit will function as it is intended to. 
Complexity of determining the key inputs will increase exponentially 
with the number of key-gates when attacker performs brute-force search. 

\vspace{-0.5em}

\subsection{SAT Attack}
Despite logic locking being successful in securing the 
IC from reverse engineering, 
the Boolean Satisfiability-based attack commonly known as SAT-attack \cite{Subramanyan_HOST'15} proposed in 2015 has successfully broke six state-of-the-art logic-locking defense mechanism proposed. 
The results have shown that the circuits can be successfully 
deobfuscated despite deploying logic locking solutions within few seconds. 
To mitigate this SAT-based attack(s), researchers have proposed several counter-measures from time to time and new attack model(s) has also been proposed to counterfeit that defense. 
We present a glimpse of SAT-attack methodology here: 

\subsubsection{Attack Model}
The attack model was established under the assumption that the attacker has
\begin{itemize}
    \item A gate-level netlist extracted from the obfuscated IC. 
    \item An activated functional chip for observing the output pattern for a given input. 
\end{itemize}

\subsubsection{Attack Methodology}
SAT attack generates a carefully crafted input patterns and observed the corresponding output from the activated functional chip. The goal of SAT attack is to eliminate incorrect key-values at each iteration by observing the outputs for a given pair of inputs. 
This input/output pairs are called Discriminating Input Patterns (DIP). By observing this DIP, SAT-attack iteratively eliminates  numerous wrong keys and this step is iterated until it eliminates all the wrong keys and determines the correct key \cite{Subramanyan_HOST'15}. 

\begin{figure*}[htbp]
\centerline{\includegraphics[width=\linewidth,height=.47\linewidth]{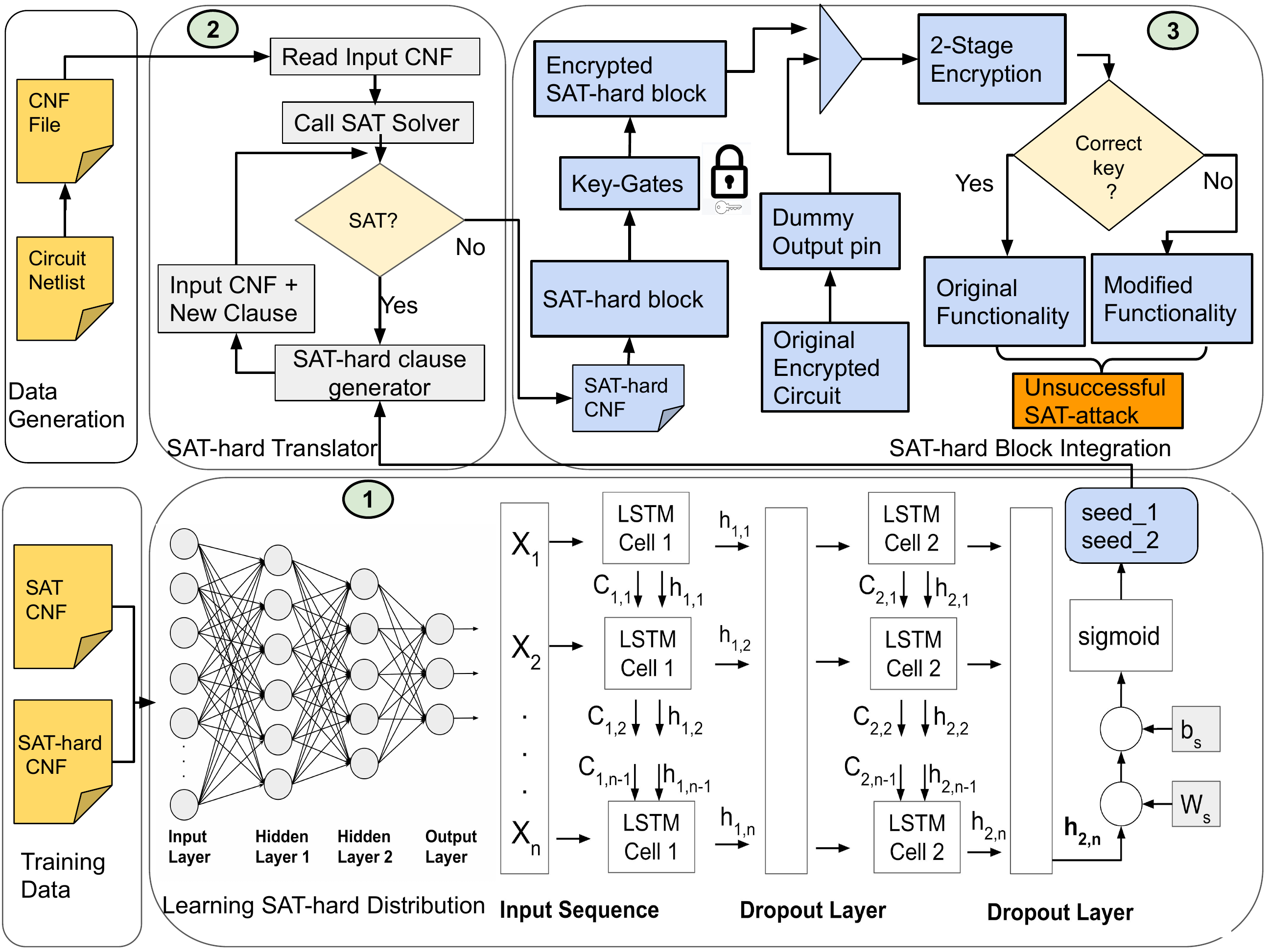}} 
\caption{Architecture of proposed SATConda. }
\label{fig:SATConda-architecture}
\end{figure*} 
\section{SATConda: SAT to UNSAT Translator}

\section{Related Work}
Several techniques have been proposed to defend and secure the design 
from SAT-attack. Here, we review some of the relevant prominent 
defense techniques to thwart SAT attacks and its variants. 


EPIC \cite{Roy_CS'10} is one of the preliminary notable work
that proposed inserting XOR/XNOR gates randomly as key-gates to 
the original netlist to achieve a logic locked netlist. 
One can decrypt the key-values by inspecting the XOR/XNOR gates and configuring them as buffers or inverters using the key-inputs\cite{Yasin_TCAS'15}.  

Insertion of an additional circuit block, Anti-SAT block \cite{Xie_CHES'16} was proposed to add with the encrypted circuit to mitigate the SAT attack. The authors showed that the time required to expose all the key-values is an exponential function of the key gates in the Anti-SAT block. By making the key-size large enough, the SAT attack becomes computationally complex and infeasible. 

Similar work has been reported in SARLock \cite{Yasin_HOST'16}. In this work they proposed a SARLock block with the encypted circuit that maximizes the number of DIPs, thus making the SAT-attack runtime exponential 
with the number of secret-key bits. This method was shown to be vulnerable to a SAT-based attack, double-DIP attack reported in \cite{Shen_GLSVLSI'17}.  

In another work, advanced encryption standard (AES) circuit \cite{Yasin_TCAS'15} was proposed into an encrypted circuit to prevent the SAT-attack. By adding this AES circuit, \cite{Yasin_TCAS'15} makes the attack computationally intractable as the attacker cannot retrieve the input of the AES block by observing the output patterns. However, this techniques suffers from 
large area overhead since implementation of AES circuit requires for significant number of logic gates \cite{Xie_CHES'16}.

In \cite{Shamsi_GLSVLSI'17} a SAT-resilient cyclic obfuscated circuit design was proposed by adding dummy paths to the encrypted circuit which make the combinational loop non-reducible. This defense mechanism was prone to another type of SAT-based attack named CycSAT\cite{Zhou_ICCAD'17}
which can effectively decrypt the cyclic encryption.

\rakibtc{
The CycSAT attack \cite{Zhou_ICCAD'17} performs a pre-processing step so that it generates a cycle avoidance clause list. The more clauses that is responsible for the SAT solver to be trapped in an infinite loop the more pre-processing is done by the CycSAT. In \cite{roshanisefat2018srclock} a number on methods are proposed to increase the cycles in a netlist exponentially to defense against the CycSAT attack. They proposed a technique to insert `Super Cycles' block that causes an exponential increase in the cycles in a given netlist and leads to a exponential increase in the runtime of the CycSAT pre-processing step. As a result, the SAT solver fails to solve the netlist within a feasible amount of time.
} 

\rakibtc{
Another defense mechanism called delay locking introduces functional key gates as well as delay key gates to a netlist. In this technique the original circuit functionality and the pre-defined timing constraints are locked using key-gates and delay-gates, respectievely \cite{xie2017delay}. The original functionality of the netlist is dependent on the correct key inputs and the correct timing constraints are satisfied when proper delay-keys are provided. 
}An in-depth review of SAT attacks and defenses is presented in \cite{hadi2019threats}.  

\tcrevision{Numerous SAT attack resilient locking techniques such as Anti-SAT \cite{Yasin_ASP-DAC_17}, SARLock \cite{roshanisefat2018srclock} have been proposed recently. In the Anti-SAT \cite{Yasin_ASP-DAC_17} technique, two complementary logic blocks are introduced and ANDed together that produce incorrect output for an incorrect key and corrupt an internal node. The SARLock mechanism \cite{roshanisefat2018srclock} stores the correct key inside a tamper-proof memory and masks the output inversion for the correct key. Both the above methods use one-point functions to achieve the resiliency against the SAT-attack. In this work, we introduced a neural network-based unSAT clause generator to integrate that block to enhance the security of an existing encryption algorithm such as XOR-based logic locking, LUT-based logic locking \cite{kolhe2019security} etc.}

Unlike the existing defenses discussed above, 
our proposed methodology utilizes a cognitive approach by utilizing neural network and extracts the feature variables automatically to learn the SAT and unSAT distributions. These distributions will be further utilized to translate a CNF from SAT to unSAT by adding additional circuitry with least overhead yet providing security compared to existing defenses. Finally, the introduced unSAT block will be encrypted with the previously SAT-attack prone obfuscated circuit to further strengthen the obfuscation and enhance resilience against other kinds of attacks. 



Our proposed defense methodology against SAT attacks is presented here. 
SATConda generates an unSAT block and encrypts the original obfuscated circuit with that block and outputs a stronger obfuscated circuit. 
In order to cognitively generate the unSAT block, 
SATConda is equipped with a hybrid Message Passing Neural Network (MPNN) \cite{Gilmer_ICML'17} framework that learns the SAT and unSAT distribution from a given number of SAT problems. 
Figure \ref{fig:SATConda-architecture} depicts the operational flow of the proposed framework and integration of the unSAT block with the obfuscated circuit, respectively. 
\vspace{-0.5em}

\subsection{SAT to unSAT Translator}

To achieve an unbreakable obfuscated circuit SATConda generates a unSAT block and integrates or modifies the existing CNF to make the whole design unSAT. As random generation of unSAT clauses is not efficient, this work 
proposes a cognitive way of generating the unSAT clauses by learning the underlying CNF pattern. 
This is performed cognitively with the aid of neural network from which the seeds for clause generation is extracted. 
The details of the neural network are discussed in the next subsections. 

The unSAT block generation process starts with getting the seed values and passing them to $seed_1$ and $seed_2$ variables (see Figure \ref{fig:SATConda-architecture}). 
The values of $seed_1$ and $seed_2$ are determined by fitting the 
learnt distribution to Bernoulli and Geometric distributions, respectively.
\rakibtc{A randomly generated decimal number between 0 and 1 is then compared with the $seed_1$ value (see Algorithm \ref{alg:satconda} line 5). Another variable $Literal\_base$ is assigned as integer 1 if the decimal number is greater than the $seed_1$ otherwise integer 2 (see Algorithm \ref{alg:satconda} line 6-8). Then the generator draws samples from a geometric distribution at a probability of $seed_2$ and assigns that value to a variable ($Rand\_geo$) (see Algorithm \ref{alg:satconda} line 10). Adding variables $Literal\_base $ and $ Rand\_geo$ we have the length of a new clause (how many literals in the clause) (see Algorithm \ref{alg:satconda} line 12).
Then SATConda samples a variable from the original CNF variable list without replacement. SATConda samples the variable with a 50\% probability of taking that variable or negating that variable.
Appending the new clause with the previous clauses SATConda again checks for the satisfiability and keeps adding clauses until the new obfuscated CNF becomes unSAT. 
\\
A function named $GENERATE\_CLAUSE$ generates each clause using the learnt parameters ( discussed in Section \ref{learning} ) and returns an unSAT block. Algorithm \ref{alg:satconda} line 23-33 shows this function. Then we integrate that block to the original circuit. We discuss this integration process in Section \ref{satconda_integration}.
\\
The range for the $seed_1$ and $seed_2$ is between 0 and 1. The greater the value of the seed the less number of clauses are required to make a circuit SAT-hard.
Algorithm \ref{alg:satconda} presents the aforementioned clause generation steps. 
}
\tcrevision{The significance of $seed_1$ and $seed_2$ is that they determine the 
achievable PPA overheads. The values of $seed_1$ and $seed_2$ are determined by fitting the learned MPNN data distribution to Bernoulli and Geometric distributions, respectively. If the learned distribution fits well to the Bernoulli and Geometric distribution then the value of $seed_1$ and $seed_2$ approaches to 1. Thus the PPA overhead is low. On the other hand if the learned distribution does not fit the Bernoulli and Geometric distribution well then the value of $seed_1$ and $seed_2$ is close to 0. This leads to a high PPA overhead.}

\subsection{unSAT Block Integration} \label{satconda_integration}
\rakib{Though the generated unSAT block can be effective against traditional SAT attacks, it can be vulnerable to additional attacks such as partitioning-based removal attacks \cite{yasin2017removal}. To address such vulnerabilities, 
we tightly integrate the generated unSAT clause (circuit) effectively with the original circuit. 
}
\rakib{Thus, our main objective is to hide the original output for a particular output pin from the attacker. To do so, SATConda adds a dummy output pin by replacing one of the original output pin, as shown in Figure \ref{fig:SATConda-architecture} (top-right block). The rationale behind introducing a dummy output pin is that the SAT-attack algorithm matches the output with the activated chip and when it fails to get the value for one output pin then the algorithm fails to decode the keys. The proposed SATConda generates a lightweight unSAT block and encrypts that block with a key-gate (AND gate). This key-gate and the dummy output (which has the original circuit's output) are XOR'ed together. The output of this XOR gate is the final output pin that was replaced with the dummy output pin. Now this output pin is encrypted with an unSAT block and a key-gate. This is a two-step encryption. When the SAT-attack algorithm tries to figure out the output of this encrypted pin it fails to do so because the SAT-solver could not solve that unSAT block. Thus, the SAT-attack algorithm fails to decrypt the value of the key-gates.
The functionality of the unSAT block is hidden and also encrypted with a key-gate, only correct key will expose this block. So, the partitioning-based removal
attack \cite{yasin2017removal} will not succeed to identify the unSAT block.
}

\tcrevision{Though the generated SAT-hard block can be effective against traditional SAT attacks such as \cite{Subramanyan_HOST'15}, one can argue that it can be vulnerable to additional attacks such as partitioning-based removal attacks \cite{yasin2017removal}. To Address such vulnerabilities, we integrate the generated SAT-hard block by hiding the outputs with the original circuit. Here, our main objective is to hide one of the original output pins from the attacker and integrate that hidden output pin with the SAT-hard block thereby thwarting such attacks. Finally, with proper key value, the original functionality of the hidden output pin is retrieved.}
\rakib{
\subsection{Learning the SAT and unSAT Distributions} \label{learning} 
\tcrevision{
As aforementioned, the proposed technique introduces a SAT-hard block and integrates it with the input CNF. The generation of the SAT-hard block is performed in a cognitive manner to alleviate large overheads. For this purpose, we utilize a message passing neural network (MPNN) similar to \cite{Selsam_CS-AI'18} in this work. The MPNN is widely studied in recent times for generic SAT problems and is yet to be  explored in the context of hardware security. The motivation for using a neural network is its ability to learn a SAT distribution and a SAT-hard distribution from the given samples. The best fit MPNN model shows the high accuracy for distinguishing a SAT problem from a SAT-hard one \cite{Selsam_CS-AI'18}. }
There are two advantages of deploying a neural network in logic locking regime. First, with the increased computation power and advancements in the SAT algorithms, the state-of-the-art SAT attacks are very advanced and powerful. 
Secondly, to protect the existing locking techniques from the powerful SAT-attacks, a cognitive counter measure is effective. 
}

The deployed MPNN encompasses of 
three layer fully-connected layers followed 
by a two-layer long short term memory (LSTM) network.
In order to learn the SAT and unSAT distributions, we encode the IC obfuscation problem as a SAT 
problem and then represent that problem as a directed 
graph where each clause and each literal is represented 
as a node individually whose dependencies will be learned through message passing.

Message is passed back and forth along the edges of 
the network \cite{Gilmer_ICML'17}. The message passing 
starts by passing a message to a clause from its neighbouring literals. In the next step, a literal gets message from its neighbouring 
clause(s) and also from its complements. This message passing event occurs back and forth until the model refines a vector space 
for every node.

The literal vector ($L_{init}$) and clause vector  ($C_{init}$) are fed to a three-layer fully connected MPNN. The output from the MPNN, $(L_{msg},C_{msg},L_{sat})$, are fed to a two-layer long-short term memory (LSTM) ($C_u, L_u$) network. Hidden states for literals and clauses are denoted by $L_h$ and $C_h$, respectively. An adjacency Matrix ($M$) defines the relationship between literals and clauses. This relationship between literals and clauses are established by connecting edges among them. The LSTM network \cite{Lei_STAT-ML'16} updates the literals $L^{t+1}$ and clauses $C^{t+1}$ at each iteration, as follows: 
\vspace{-0.5em}
\begin{equation}
     C_u([M^TL_{msg}(L^t)]) \rightarrow C^{t+1}
\end{equation}
\begin{equation}
    C_u([C_h^t]) \rightarrow C^{t+1}_h
\end{equation}
\begin{equation}
     L_u([Flip(L^t), MC_{msg}(C^{t+1})]) \rightarrow L^{t+1}
\end{equation}
\begin{equation}
    L_u([L_h^t]) \rightarrow L^{t+1}_h
\end{equation}

$L_{sat}$ votes SAT or unSAT for a particular literal and taking the average vote of all the literals after $T$ iteration, SATConda predicts whether a problem is SAT or unSAT.
This message-passing architecture lets the SATConda to learn the features that can distinguish the SAT solvable CNFs from unSAT CNFs i.e., learn the distribution of SAT and unSAT CNFs.
In order to generate the unSAT clause, we choose the seed values that 
best-fit unSAT distribution and utilize it for unSAT clause generator to generate an unSAT block with minimum number of clauses yet secure. 
In other words, the $seed_1$ and $seed_2$ values are determined as the values that best fit the unSAT distribution. 
This process leads to minimum area and power overhead which is one of the main challenges for logic locking and other obfuscation techniques.

\tcrevision{
\subsection{Training and Testing The MPNN Model}
To train the neural network (NN) for learning SAT and SAT-hard distributions we provide the NN a pair of SAT problems. In this pair, one problem is satisfiable (SAT) and the other one is SAT-hard. In the training pool, we have a total of $m$ SAT problems and $n$ SAT-hard problems where $\mathcal{C}_m$ = $\{C_1, C_2,\cdots,C_m\}$ and $\mathcal{C}_n$ = $\{C_1', C_2',\cdots,C_n\}$. Each of the SAT and SAT-hard problems 
has $k$ number of clauses with each clause having $n$ variables (ranges from 5 to 20). In some cases, the difference between SAT and SAT-hard samples could be a simple flipping of a literal in a clause. 
} 

\tcad{
Each of the training data is labeled either 0 or 1, where 0 means SAT-hard and 1 stands for satisfiable (SAT). For a \{${SAT},{SAT-hard}$\} pair, the fully connected MLP tries to update the weights for each neuron and learn the relationship between the feature vectors, i.e., clauses and literals. The LSTM network is then updated with the predicted output from the MLP and continue to iterate the process. After a number of iterations the network predicts whether a problem is SAT or SAT-hard. We trained our MPNN model on ten thousand SAT problems (out of which five thousand are SAT-hard) for better learning and efficient modeling of SAT and SAT-hard distributions. The model that achieves the highest accuracy  is chosen to generate the SAT-hard block in this work. For the best-fit model, the training-cost is 0.6930 and the validation cost is 0.6932, which is sufficiently good enough to ensure that the model generalizes well with no over or under-fitting.}


\begin{algorithm}[htb!]\small
\caption{SATConda Algorithm}\label{alg:satconda}
$\textbf{Input}: solve\_clauses, seed\_1, seed\_2 \\$
$\textbf{Output}: unSAT-cnf$
\begin{algorithmic}[1]
\State $is\_sat := solve(solve\_clauses)$
    \If{$is\_sat == True$}
      \While{true}
      \State $rand := gen\_decimal(0-1)$
        \If{$rand < seed_1$}
            \State $lit\_base := 1$
        \Else
            \State $lit\_base := 2$
        \EndIf
        \State $rand\_geo = rand.geometric(seed\_2)$
        \State $literal = lit\_base + rand\_geo $
        \State $new\_clause := generate\_clause(n\_var, literal)$
        \State $solve\_clauses += new\_clause $
        \State $is\_sat := solve(solve\_clauses)$
        \If {$is\_sat == True$}
          \State $solve\_clauses += new\_clause$
        \Else
          \State $break$
        \EndIf
      \EndWhile
      \State $solve\_clauses += new\_clause$
    \EndIf
\Function{generate\_clause}{$n\_var$, $literal$}
    \State $array\_size := minimum(n\_var, literal)$
    \State $clause\_gen := gen.rand\_array(n\_var, array\_size)$
    \State $rand := gen\_decimal(0~1)$
    \If {$rand < 0.5$}
    \State $new\_clause := clause\_gen + 1$
    \Else
    \State $new\_clause := -(clause\_gen + 1)$
    \EndIf
    \State \textbf{return} $new\_clause$
\EndFunction
\State $unSAT-cnf := solve\_clauses$
\end{algorithmic}
\end{algorithm}

\section{Experimental Results}
In this section we describe the experimental setup  
 and evaluate the impact of SATConda in terms of SAT hardness and the incurred overheads.  

\subsection{Experimental Setup}
In this work, we used the MPNN model similar to \cite{Selsam_CS-AI'18} 
to obtain the seed required for our random clause generator. We trained 
the model with 10,000 CNFs, out of which 5000 are SAT and 5000 are unSAT.

\tcrevision{We evaluate the performance on ISCAS'85 and ISCAS'89 benchmark circuits shown in Table \ref{tab:isca-bench}.
We used two different obfuscation techniques to encrypt ISCAS'85 and ISCAS'89 
benchmark circuits in this work. For ISCAS'85 benchmark, one algorithm was proposed by Rajendran et al. \cite{DAC_12} that inserts XOR/XNOR gates (referred to as ``DAC'12") at different locations of the circuit to prevent the fault-analysis attack. The second algorithm was proposed by Dupuis et al. \cite{IOLTS_14} that inserts AND/OR gates (referred to as ``IOLTS'14") at multiple chosen locations of the circuit. }

\tcrevision{For ISCAS'89 benchmark, we used the IOLTS'14 obfuscation \cite{IOLTS_14} technique. 
It needs to be noted that the algorithm \cite{Subramanyan_HOST'15} was unable 
to encrypt the ISACAS'89 circuit using DAC'12 algorithm despite executing it for three days. Instead of DAC'12, we applied a similar technique that is 
also a XOR-based logic locking (referred to as ``TOC'13") algorithm \cite{toc13xor} to obfuscate ISCAS'89 benchmark circuit, which is also proposed by the same group. It needs to noted that the proposed technique enhances the security of the CNF and is independent of the underlying obfuscation technique. We considered these obfuscation techniques as mere case studies to show the effectiveness. 
}

\tcrevision{While generating the  obfuscated benchmark circuits for XOR/XNOR-based logic locking \cite{DAC_12}, AND/OR-based logic locking \cite{IOLTS_14}, and XOR-based logic locking \cite{toc13xor}, we considered 5\% as the limit on number of gates to be obfuscated. 
As the circuit size increases, the number of key-bit as well increases. Thus, depending on the original circuit area, the number of key bits vary. The number of key bits ranges from 6 (for the smallest circuit) to 186 (for the largest circuit). On top of the obfuscated benchmark circuit through aforementioned techniques, the proposed technique needs one more key bit to integrate the unSAT block to the original obfuscated circuit.}
\tcrevision{The number of iterations required for the clause generation process depend on various factors. One key factor being the learnt parameters, i.e., seed values, from the neural network. We have discussed the significance of seed values in Section 4.2. In summary, if the learned distribution fits well to the Bernoulli and Geometric distributions, 
then the value of $seed_1$ and $seed_2$ approaches to 1. Thus, the algorithm requires less number of iterations for the clause generation process. For our best-trained model, the number of iterations ranges from 60 iterations (for the smallest circuits) to 1005 iterations.}

In addition to verification against traditional SAT attacks used in hardware security domain, we have also verified the satisfiability of CNF using three different SAT solvers, MiniSAT \cite{Sorensson_SAT'05}, Lingeling \cite{Biere_SAT'13}, and Glucose \cite{Audemard_SAT'09}. The rationale for choosing these solvers is that these solvers form basis for numerous
SAT attacks crafted for deobfuscation in the past few years. The area and power overheads are calculated using Synopsys Design Compiler, Version: L-2016.03-SP3. SAED 90nm EDK Digital Standard Cell Library \cite{Goldman_MSE'09} is used for logic synthesis. All the experiments were performed on a server with 8-core Intel Xeon E5410 CPU, running CentOS Linux 7 at 2.33 GHz, with 16 GB RAM. 
\vspace{-0.5em}
\subsection{Evaluation}

Here, we present the evaluation in terms of SAT-hardness and the overhead analysis in addition to our empirical findings of SATConda. 
\subsubsection{SAT-hardness}
\rakib{
Table \ref{tab:sat-attack-check} shows SAT-attack \cite{Subramanyan_HOST'15} performance on two different encryption algorithms \cite{IOLTS_14, DAC_12} on ISCAS'85 benchmarks before and after they are passed through SATConda. It shows that the SAT-attack in \cite{Subramanyan_HOST'15} successfully decrypts the keys of the obfuscated circuit (obfuscated by \cite{IOLTS_14},\cite{DAC_12}) in less than a minute. However, the same SAT-attack fails to extract the keys when it is further encrypted using SATConda. For both the XOR/XNOR-based logic locking \cite{DAC_12}  and AND/OR-based logic locking \cite{IOLTS_14}, SATConda successfully defends the SAT-attack for all the circuits. The timeout for the SAT-attack is set to 10 hours in our experiments. In case of the c2670 circuit, encryption in \cite{DAC_12} itself resists against SAT attack, thus, SATConda does not introduce additional circuit.
Table \ref{tab:sat_isca89} presents the satisfiability of the ISCAS'89 benchmark circuits on IOLTS'14 and TOC'13 encryption before and after the conversion through SATConda.
From the results it could be observed that the proposed SATConda can enhance the security of the underlying obfuscation techniques. }

\tcrevision{
In addition to SAT attack in \cite{Subramanyan_HOST'15}, we have also evaluated the robustness with SATConda in terms of SAT-hardness against other SAT attacks such as AppSAT \cite{shamsi2017appsat}. Table \ref{tab:appsat_isca85} represents the SATConda  evaluation  on  AppSAT-attack \tcrevision{\cite{shamsi2017appsat}} for  ISCAS’85 benchmark circuits.  As can be observed that the proposed SATConda is able to resist the AppSAT attack, which earlier was able to decrypt the benchmark circuits. 
} 
\tcrevision{
Table \ref{tab:appsat_isca89} represents the SATConda  evaluation  on  AppSAT attack  for  ISCAS’89 benchmark circuits. 
Similar resiliency against AppSAT attack is observed with proposed SATConda when applied on both encryption schemes. 
} 

As it is nearly impossible to evaluate against all the existing and future SAT attacks, we consider the traditional SAT attacks which form basis for most of the present day SAT-based deobfuscation attacks for evaluation. 
The satisfiability of the encryption techniques were verified against three different SAT solvers, MiniSAT \cite{Sorensson_SAT'05}, Lingeling \cite{Biere_SAT'13}, and Glucose \cite{Audemard_SAT'09}. 
Table \ref{tab:sat-unsat-check} presents the satisfiability of the ISCAS'85 benchmark circuits on IOLTS'14 encryption before and after the conversion through SATConda. It can be seen 
that all the encrypted benchmark circuits using IOLTS'14 encryption were satisfiable (breakable) with all the three traditional SAT-solvers \cite{Sorensson_SAT'05,Biere_SAT'13,Audemard_SAT'09}. 
This indicates that an attacker could perform a SAT-attack with any of these SAT-solvers and reverse engineer the IP/IC. Table \ref{tab:sat-unsat-check} also shows that once the IC/IP design when translated using SATConda, it becomes SAT-hard, 
indicating that none of the three experimented SAT-solvers, which were previously successful, could solve a satisfying assignment.

\tcrevision{
Table \ref{tab:sat-unsat-check-dac} shows the satisfiability of the same ISCAS'85 benchmark circuits on DAC'12 logic locking \cite{DAC_12} before and after the conversion through SATConda. Similar to the IOLTS'14 logic locking \cite{IOLTS_14}, the DAC'12 logic locking \cite{DAC_12} was vulnerable to these SAT solvers.  When this encrypted circuit was further passed through SATConda it becomes SAT-hard. Meaning that with the specified timeout window the SAT solvers fails to satisfy for any assignment.
Table \ref{tab:sat-unsat-check-iolts89} and \ref{tab:sat-unsat-check-toc89} present the satisfiability result for ISCAS'89 circuits using the IOLTS'14)logic locking  \cite{IOLTS_14} and TOC'13 logic locking  \cite{toc13xor} methods, respectively. 
Similar to ISCAS'85 benchmarks, ISCAS'89 benchmarks also showcase resiliency against the traditional SAT attacks when passed through the proposed SATConda. } 


\begin{table}[htbp]
  \centering
  \caption{ISCAS'85 and ISCAS'89 benchmark circuits}
  \label{tab:isca-bench}
  \begin{tabularx}{\columnwidth}{|X|X|X|X|X|X|X|X|} \hline
  \multicolumn{4}{|c|}{ISCAS'85 Circuit} & \multicolumn{4}{c|}{ISCAS'89 Circuit} \\ \hline
    Circuit Name & \#Inputs & \#Outputs & \#Gates & Circuit Name & \#Inputs & \#Outputs & \#Gates \\ \hline
 &  &  &  & s382 & 24 & 27 & 392 \\ \hline 
c432 & 36 & 7 & 160 &  s400 & 24 & 27 & 414\\ \hline  
c499 & 41 & 32 & 202 & s641 & 54 & 42 & 459 \\ \hline 
c880 & 60 & 26 & 383 & s526n & 24 & 27 & 494 \\ \hline 
c1355 & 41 & 32 & 546 & s526 & 3 & 6  & 141 \\ \hline 
c1908 & 33 & 25 & 880 & s953 & 45 & 29 & 950 \\ \hline 
c2670 & 233 & 140 & 1193 & s1488 & 14 & 25 & 843 \\ \hline  
c3540 & 50 & 22 & 1669 & s5378 & 214 & 228 & 5183 \\ \hline 
c5315 & 178 & 123 & 2307 & s13207 & 700 & 790 & 11248 \\ \hline  
c7552 & 207 & 108 & 3512 & s15850 & 611 & 684 & 13192 \\ \hline
 & & & & s35932 & 1763 & 1728 & 31833 \\ \hline
    \end{tabularx}
    \end{table}
    
\begin{table}[htbp]
    \centering
    \caption{SATConda evaluation on SAT-attack \cite{Subramanyan_HOST'15} for ISCAS'85 circuit with different encryption techniques}
    \label{tab:sat-attack-check}
    \begin{tabularx}{\columnwidth}{|X|X|X|X|X|} \hline
    & \multicolumn{2}{c|}{IOLTS'14 \cite{IOLTS_14}} & \multicolumn{2}{c|}{DAC'12 \cite{DAC_12}} \\ 
\cline{2-5}
    & \multicolumn{2}{c|}{Time (s)} & \multicolumn{2}{c|}{Time (s)} \\ 
\cline{2-5}
 Circuit Name  & Before Conversion  & After Conversion & Before Conversion & After Conversion \\ \hline
c432 & 0.033 & timeout & 0.026 & timeout  \\ \hline
c499 & 0.060 & timeout & 0.070 & timeout  \\ \hline
c880 & 0.061 & timeout & 0.094 & timeout \\ \hline
c1355 & 0.042 & timeout & 0.312 & timeout  \\ \hline
c1908 & 0.049 & timeout & 0.518 & timeout  \\ \hline
c2670 & 0.544 & timeout & timeout & **  \\ \hline
c3540 & 0.323 & timeout & 3.264 & \rakib{timeout}  \\ \hline
c5315 & 0.826 & timeout & 9.013 & \rakib{timeout}  \\ \hline
c7552 & 0.467 & timeout & 26.75 & \rakib{timeout}  \\ \hline
\multicolumn{5}{l}{timeout = 10 hours} \\
\multicolumn{5}{l}{** = Excluded from the experiment as the original encrypted circuit} \\
\multicolumn{5}{l}{was unbreakable by SAT-attack} \\
    \end{tabularx}
    \end{table}

\begin{table}[htbp]
    \centering
    \caption{SATConda evaluation on SAT-attack \cite{Subramanyan_HOST'15} for ISCAS'89 circuits with different encryption techniques}
    \tcrevision{
    \label{tab:sat_isca89}
    \begin{tabularx}{\columnwidth}{|X|X|X|X|X|} \hline
    & \multicolumn{2}{c|}{IOLTS'14 \cite{IOLTS_14}} & \multicolumn{2}{c|}{ToC'13 \cite{toc13xor}} \\ 
\cline{2-5}
    & \multicolumn{2}{c|}{Time (s)} & \multicolumn{2}{c|}{Time (s)} \\ 
\cline{2-5}
 Circuit Name  & Before Conversion  & After Conversion & Before Conversion & After Conversion \\ \hline
s382 & 0.006 & timeout & 0.006 & timeout  \\ \hline
s400 & 0.006 & timeout & 0.006 & timeout  \\ \hline
s526n & 0.0087 & timeout & 0.008 & timeout \\ \hline
s526 & 0.0084 & timeout & 0.008 & timeout  \\ \hline
s641 & 0.009 & timeout & 0.0092 & timeout  \\ \hline
s953 &  0.017 & timeout &  0.017 & timeout  \\ \hline
s1488 & 0.048 & timeout & 0.0344 & timeout  \\ \hline
s5378 & 0.070  & timeout & 0.072 & timeout  \\ \hline
s13207 & 0.252 & timeout & 0.232 & timeout  \\ \hline
s15850 & 0.328 & timeout & 0.336 & timeout  \\ \hline
s35932 & 5.831 & timeout & 5.890 & timeout  \\ \hline
\multicolumn{5}{l}{timeout = 10 hours} \\
    \end{tabularx}
    }
    \end{table}

\begin{table}[htbp]
    \centering
    \caption{SATConda evaluation on AppSAT-attack \cite{shamsi2017appsat} for ISCAS'85 circuits}
    \label{tab:appsat_isca85}
    \tcrevision{
    \begin{tabularx}{\columnwidth}{|X|X|X|X|X|} \hline
    & \multicolumn{2}{c|}{IOLTS'14 \cite{IOLTS_14}} & \multicolumn{2}{c|}{ToC'13 \cite{toc13xor}} \\ 
\cline{2-5}
    & \multicolumn{2}{c|}{Time (s)} & \multicolumn{2}{c|}{Time (s)} \\ 
\cline{2-5}
 Circuit Name  & Before Conversion  & After Conversion & Before Conversion & After Conversion \\ \hline
c432 & 0.055 & timeout & 0.058 & timeout  \\ \hline
c499 & 0.085 & timeout & 0.086 & timeout  \\ \hline
c880 & 0.1598 & timeout & 0.20 & timeout \\ \hline
c1355 & 0.098 & timeout & 0.219 & timeout  \\ \hline
c1908 & 0.122 & timeout & 3.945 & timeout  \\ \hline
c2670 & 1.712 & timeout & 1.380 & timeout  \\ \hline
c3540 & 1.968 & timeout & 4.921 & timeout  \\ \hline
c5315 & 2.888 & timeout & 1.71 & timeout  \\ \hline
c7552 & 4.506 & timeout & 2.94 & timeout  \\ \hline
\multicolumn{5}{l}{timeout = 10 hours} \\
    \end{tabularx}
    }
    \end{table}

\begin{table}[htbp]
    \centering
    \caption{SATConda evaluation on AppSAT-attack \cite{shamsi2017appsat} for ISCAS'89 circuits with different encryption schemes}
    \label{tab:appsat_isca89}
    \tcrevision{
    \begin{tabularx}{\columnwidth}{|X|X|X|X|X|} \hline
    & \multicolumn{2}{c|}{IOLTS'14 \cite{IOLTS_14}} & \multicolumn{2}{c|}{ToC'13 \cite{toc13xor}} \\ 
\cline{2-5}
    & \multicolumn{2}{c|}{Time (s)} & \multicolumn{2}{c|}{Time (s)} \\ 
\cline{2-5}
 Circuit Name  & Before Conversion  & After Conversion & Before Conversion & After Conversion \\ \hline
s382 & 0.066 & timeout &  0.032 & timeout  \\ \hline
s400 & 0.067 & timeout &  0.033 & timeout  \\ \hline
s526n & 0.089 & timeout & 0.044 & timeout \\ \hline
s526 & 0.090 & timeout & 0.045 & timeout  \\ \hline
s641 & 0.112 & timeout & 0.053 & timeout  \\ \hline
s953 &  0.157 & timeout &  0.077 & timeout  \\ \hline
s1488 & 0.316 & timeout & 0.117 & timeout  \\ \hline
s5378 &  0.541 & timeout & 0.268 & timeout  \\ \hline
s13207 & 1.52 & timeout & 0.588 & timeout  \\ \hline
s15850 &  2.265 & timeout & 0.887 & timeout  \\ \hline
s35932 & 17.882 & timeout & 8.138 & timeout  \\ \hline
\multicolumn{5}{l}{timeout = 10 hours} \\
    \end{tabularx}
    }
    \end{table}

\begin{table}[htbp]
    \centering
    \caption{SATConda evaluation using IOLTS'14 \cite{IOLTS_14} encryption on different SAT-solvers for ISCAS'85 benchmarks}
    \label{tab:sat-unsat-check}
    \begin{tabularx}{\columnwidth}{|X|X|X|X|X|X|X|} \hline
    & \multicolumn{2}{c|}{miniSAT \cite{Sorensson_SAT'05}} & \multicolumn{2}{c|}{Lingeling \cite{Biere_SAT'13}} & \multicolumn{2}{c|}{Glucose \cite{Audemard_SAT'09}}  \\ 
\cline{2-7}
        & \multicolumn{2}{c|}{Time(s)} & \multicolumn{2}{c|}{Time(s)} & \multicolumn{2}{c|}{Time(s)} \\ 
\cline{2-7}
 Circuit Name  & Before Conversion & After Conversion & Before Conversion & After Conversion & Before Conversion & After Conversion  \\ \hline
c432 & 0.0039 & \xmark & 0.1 & \xmark & 0.0015 & \xmark   \\ \hline
c499 & 0.0026 & \xmark & 0.1 & \xmark & 0.0011 & \xmark    \\ \hline
c880 & 0.0024 & \xmark & 0.1 & \xmark & 0.0014 & \xmark    \\ \hline
c1355 & 0.0018 & \xmark & 0.1 & \xmark & 0.0005 & \xmark    \\ \hline
c1908 & 0.0041 & \xmark & 0.1 & \xmark & 0.0031 & \xmark    \\ \hline
c2670 & 0.0049 & \xmark & 0.1 & \xmark & 0.0003 & \xmark    \\ \hline
c3540 & 0.0060 & \xmark & 0.1 & \xmark & 0.0061 & \xmark    \\ \hline
c5315 & 0.0086 & \xmark & 0.1 & \xmark & 0.0094 & \xmark    \\ \hline
c7552 & 0.0095 & \xmark & 0.1 & \xmark & 0.0110 & \xmark    \\ \hline
\multicolumn{7}{l}{\xmark  =  Corresponding CNF was unsatisfiable by the SAT solver }
    
    \end{tabularx}
    \end{table}
    
\begin{table}[htbp]
    \centering
    \caption{SATConda evaluation using DAC'12 encryption \cite{DAC_12} on different SAT-solvers for ISCAS'85 benchmarks}
    \label{tab:sat-unsat-check-dac}
    \begin{tabularx}{\columnwidth}{|X|X|X|X|X|X|X|} \hline
    & \multicolumn{2}{c|}{miniSAT \cite{Sorensson_SAT'05}} & \multicolumn{2}{c|}{Lingeling \cite{Biere_SAT'13}} & \multicolumn{2}{c|}{Glucose \cite{Audemard_SAT'09}}  \\ 
\cline{2-7}
        & \multicolumn{2}{c|}{Time(s)} & \multicolumn{2}{c|}{Time(s)} & \multicolumn{2}{c|}{Time(s)} \\ 
\cline{2-7}
 Circuit Name  & Before Conversion & After Conversion & Before Conversion & After Conversion & Before Conversion & After Conversion  \\ \hline
c432 & 0.0049 & \xmark & 0.1 & \xmark & 0.004249 & \xmark   \\ \hline
c499 & 0.005691 & \xmark & 0.1 & \xmark & 0.004277 & \xmark    \\ \hline
c880 & 0.006145 & \xmark & 0.1 & \xmark & 0.005559 & \xmark    \\ \hline
c1355 & 0.006179 & \xmark & 0.1 & \xmark & 0.006117 & \xmark    \\ \hline
c1908 & 0.006226 & \xmark & 0.1 & \xmark & 0.007626 & \xmark    \\ \hline
c2670 & 0.003507 & \xmark & 0.1 & \xmark & 0.008887 & \xmark    \\ \hline
c3540 & 0.001801 & \xmark & 0.1 & \xmark & 0.009222 & \xmark    \\ \hline
c5315 & 0.005401 & \xmark & 0.1 & \xmark & 0.009447 & \xmark    \\ \hline
c7552 & 0.010083 & \xmark & 0.1 & \xmark & 0.010702 & \xmark    \\ \hline
\multicolumn{7}{l}{\xmark  =  Corresponding CNF was unsatisfiable by the SAT solver }
    
    \end{tabularx}
    \end{table}

\begin{table}[htbp]
    \centering
    \caption{SATConda with IOLTS'14 encryption \cite{IOLTS_14} evaluated on different SAT-solvers for ISCAS'89 circuits}
    \label{tab:sat-unsat-check-iolts89}
    \tcrevision{
    \begin{tabularx}{\columnwidth}{|X|X|X|X|X|X|X|} \hline
    & \multicolumn{2}{c|}{miniSAT \cite{Sorensson_SAT'05}} & \multicolumn{2}{c|}{Lingeling \cite{Biere_SAT'13}} & \multicolumn{2}{c|}{Glucose \cite{Audemard_SAT'09}}  \\ 
\cline{2-7}
        & \multicolumn{2}{c|}{Time(s)} & \multicolumn{2}{c|}{Time(s)} & \multicolumn{2}{c|}{Time(s)} \\ 
\cline{2-7}
 Circuit Name  & Before Conversion & After Conversion & Before Conversion & After Conversion & Before Conversion & After Conversion  \\ \hline
s382 & 0.0016 & \xmark & 0.2 & \xmark & 0.00347 & \xmark   \\ \hline
s400 & 0.0014 & \xmark & 0.2 & \xmark &  0.0031 & \xmark    \\ \hline
s526n & 0.0015 & \xmark & 0.2 & \xmark & 0.0045 & \xmark    \\ \hline
s526 & 0.0061 & \xmark & 0.2 & \xmark & 0.0035 & \xmark    \\ \hline
s641 & 0.0014 & \xmark & 0.2 & \xmark & 0.0025 & \xmark    \\ \hline
s953 & 0.0018 & \xmark & 0.2 & \xmark & 0.0012 & \xmark    \\ \hline
s1488 & 0.0041 & \xmark & 0.2 & \xmark & 0.0070 & \xmark    \\ \hline
s5378 & 0.0034  & \xmark & 0.2 & \xmark & 0.0076 & \xmark    \\ \hline
s13207 & 0.0079 & \xmark & 0.2 & \xmark & 0.0065 & \xmark    \\ \hline
s15850 & 0.0068 & \xmark & 0.2 & \xmark & 0.0121 & \xmark    \\ \hline
s35932 & 0.0155 & \xmark & 0.2 & \xmark & 0.0202 & \xmark    \\ \hline
\multicolumn{7}{l}{\xmark  =  Corresponding CNF was unsatisfiable by the SAT solver } \\
    \end{tabularx}
    }
    \end{table}

\begin{table}[htbp]
    \centering
    \caption{SATConda evaluation using TOC'13 encryption \cite{toc13xor} on different SAT-solvers for ISCAS'89 circuits}
    \label{tab:sat-unsat-check-toc89}
   \tcrevision{
    \begin{tabularx}{\columnwidth}{|X|X|X|X|X|X|X|} \hline
    & \multicolumn{2}{c|}{miniSAT \cite{Sorensson_SAT'05}} & \multicolumn{2}{c|}{Lingeling \cite{Biere_SAT'13}} & \multicolumn{2}{c|}{Glucose \cite{Audemard_SAT'09}}  \\ 
\cline{2-7}
        & \multicolumn{2}{c|}{Time(s)} & \multicolumn{2}{c|}{Time(s)} & \multicolumn{2}{c|}{Time(s)} \\ 
\cline{2-7}
 Circuit Name  & Before Conversion & After Conversion & Before Conversion & After Conversion & Before Conversion & After Conversion  \\ \hline
s382 & 0.001 & \xmark & 0.2 & \xmark & 0.0007 & \xmark   \\ \hline
s400 & 0.001 & \xmark & 0.2 & \xmark &  0.0006 & \xmark    \\ \hline
s526n & 0.001 & \xmark & 0.2 & \xmark & 0.0007 & \xmark    \\ \hline
s526 & 0.001 & \xmark & 0.2 & \xmark & 0.0009 & \xmark    \\ \hline
s641 & 0.001 & \xmark & 0.2 & \xmark & 0.0007 & \xmark    \\ \hline
s953 & 0.002 & \xmark & 0.2 & \xmark & 0.001 & \xmark    \\ \hline
s1488 & 0.005 & \xmark & 0.2 & \xmark & 0.002 & \xmark    \\ \hline
s5378 & 0.003  & \xmark & 0.2 & \xmark & 0.002 & \xmark    \\ \hline
s13207 & 0.009 & \xmark & 0.2 & \xmark & 0.009 & \xmark    \\ \hline
s15850 & 0.009 & \xmark & 0.2 & \xmark & 0.007 & \xmark    \\ \hline
s35932 & 0.015 & \xmark & 0.2 & \xmark & 0.012 & \xmark    \\ \hline
\multicolumn{7}{l}{\xmark  =  Corresponding CNF was unsatisfiable by the SAT solver } \\
    \end{tabularx}
    }
    \end{table}

\begin{table*}[htb!]
    \caption{Area and power overhead analysis of SATConda with different encryption schemes  for ISCAS'85 circuits}
    \label{tab:overhead}
    \begin{tabularx}{\textwidth}{X||XXX|XXX|X||XXX|XXX|X}
    \hline
	&	\multicolumn{7}{c}{Area Overhead } &	\multicolumn{7}{c}{Power Overhead } 	\\ \cline{1-15}
    Circuit & IOLTS- '14 \cite{IOLTS_14} $(\mu m^2)$ & SAT- Conda+\cite{IOLTS_14}  $(\mu m^2)$ & Over- head $(\%)$ & DAC12 \cite{DAC_12} $(\mu m^2)$ & SAT- Conda+\cite{DAC_12} $(\mu m^2)$ & Over- head $(\%)$ & Over- head $(\%)$ Other Method & IOLTS- '14 \cite{IOLTS_14} $(\mu W)$ & SAT- Conda+\cite{IOLTS_14} $(\mu W)$ & Over- head $(\%)$ & DAC12 \cite{DAC_12} $(\mu W)$ & SAT- Conda+\cite{DAC_12} $(\mu W)$ & Over- head $(\%)$ & Over- head $(\%)$ Other Method \\ \hline
    c432 & 1,058.79 & 2,079.20 & 96.38 & 1,113.92 & 2,158.15 & 93.74 & 21 \cite{Yasin_TCAS'15} & 18.21 & 37.40 & 105.34 & 19.62 & 39.42 & 100.91 & 70 \cite{Yasin_TCAS'15}  \\ \hline
    c499 & 1,918.88 & 3,044.98 & 58.69 & 1,978.63 & 2,867.02 & 44.90 & -  & 51.82 & 75.17 & 45.06 & 54.60 & 72.61 & 32.98  &  - \\ \hline
    c880 & 1,613.51 & 2,861.77 & 77.36 & 1,754.96 & 2,756.61 & 57.08 &  -  & 31.19 & 53.01 & 69.97 & 38.65 & 52.62 & 36.13  & -  \\ \hline
    c1355 & 2,051.21 & 3,683.81 & 79.59 & 2,154.95 & 3,256.43 & 51.11 &  -  & 53.42 & 79.66 & 49.10 & 61.45 & 81.97 & 33.39 & -   \\ \hline
    c1908 & 2,170.10 & 3,553.32 & 63.74 & 2,628.10 & 3,587.28 & 36.50 &  -  & 49.19 & 70.46 & 43.25 & 68.53 & 78.84 & 15.05 & -   \\ \hline
    c3540 & 4,971.22 & 7,258.89 & 46.02 & 5,716.99 & 7,151.22 & 25.09 &  -  & 91.44 & 131.96 & 44.31 & 125.19 & 151.26 & 20.82 & -    \\ \hline
    c5315 & 7,279.01 & 10,770.00 & 47.96 & 8,660.90 & 12,231.05 & 41.22 & 8 \cite{Yasin_HOST'16}   & 148.50 & 198.50 & 33.67 & 205.55 & 271.21 & 31.94 & 9 \cite{Yasin_HOST'16}    \\ \hline
    c7552 & 9,697.55 & 14,856.48 & 53.20 & 10,900.76 & 17,061.17 & 56.51 & 265 \cite{Kolhe_GLSVLSI'19}   & 219.34 & 282.67 & 28.87 & 297.66 & 388.55 & 30.54 &  14 \cite{Kolhe_GLSVLSI'19}  \\ \hline
    \textbf{Average} &  &  & 57.7 &  &  & 42.8 &    &  &  & 54.44 &  &  & 36.54 &    \\ \hline
    \end{tabularx}
    \end{table*}
    
\begin{table}[htb!]
    \caption{Delay overhead analysis of SATConda with different encryption schemes  for ISCAS'85 circuits}
    \label{tab:delay-overhead}
    \begin{tabularx}{\columnwidth}{X|XXXXXX} 
    \hline
	&	\multicolumn{6}{c}{Delay Overhead } 	\\ \cline{1-7}
    Circuit & IOLTS- '14 \cite{IOLTS_14} $(ns)$ & SAT- Conda+\cite{IOLTS_14}  $(ns)$ & Over- head $(\%)$ & DAC12 \cite{DAC_12} $(ns)$ & SAT- Conda+\cite{DAC_12} $(ns)$ & Over- head $(\%)$  \\ \hline
    c432 & 1.01 & 1.36 & 34.65 & 1.12 & 1.39 & 24.11  \\ \hline
    c499 & 2.57 & 2.9 & 12.84 & 2.73 & 3.76 & 37.73    \\ \hline
    c880 & 2.9 & 3.75 & 29.31 & 3.1 & 3.41 & 10     \\ \hline
    c1355 & 2.6 & 3.44 & 32.31 & 2.71 & 3.49 & 28.78   \\ \hline
    c1908 & 3.72 & 5.01 & 34.68 & 3.52 & 4.91 & 39.49   \\ \hline
    c3540 & 4.85 & 5.58 & 15.05 & 4.84 & 5.85 & 20.87   \\ \hline
    c5315 & 3.55 & 4.73 & 33.24 & 5.22 & 7.01 & 34.29  \\ \hline
    c7552 & 4.86 & 4.99 & 2.67 & 7.05 & 7.12 & 0.99    \\ \hline
    \textbf{Average} &  &  & 24.34 &  &  & 24.53     \\ \hline
    \end{tabularx}
    \end{table}

\begin{table*}[htb!]
    \caption{Area and power overhead analysis of SATConda with different encryption schemes for ISCAS'89 circuits}
    \label{tab:overhead_89}
   \tcrevision{
    \begin{tabularx}{\textwidth}{X||XXX|XXX||XXX|XXX} 
    \hline
	&	\multicolumn{6}{c}{Area Overhead } &	\multicolumn{6}{c}{Power Overhead } 	\\ \cline{1-13}
    Circuit & IOLTS- '14 \cite{IOLTS_14} $(\mu m^2)$ & SAT- Conda+\cite{IOLTS_14}  $(\mu m^2)$ & Over- head $(\%)$ & TOC'13- XOR \cite{toc13xor} $(\mu m^2)$ & SAT- Conda+\cite{toc13xor} $(\mu m^2)$ & Over- head $(\%)$ & IOLTS- '14 \cite{IOLTS_14} $(\mu W)$ & SAT- Conda+\cite{IOLTS_14} $(\mu W)$ & Over- head $(\%)$ & TOC'13- XOR \cite{toc13xor}  $(\mu W)$ & SAT- Conda+ \cite{toc13xor}  $(\mu W)$ & Over- head $(\%)$ \\ \hline
    s382 & 662.47 & 1249.53 & 88.62 & 662.47 & 1349.04 & 103.63  & 10.31 & 20.47 & 98.52 & 10.31 & 22.92 & 122.26   \\ \hline
    s400 & 657.4 & 1328.3 & 102.05 & 657.40 & 1329.07 & 102.17 & 10.07 & 22.35 & 121.91 & 10.07 & 22.86 & 127.01   \\ \hline
    s526n & 871.94 & 1649.53 & 89.1 & 871.94 & 1618.36 & 85.6  & 13.58 & 27.27 & 100.85 & 38.65 & 52.62 & 36.13   \\ \hline
    s526 & 862.18 & 1589.89 & 84.4 & 862.18 & 1701.78 & 97.38  & 13.17 & 26.08 & 98.02 & 13.58  & 26.66  & 96.31   \\ \hline
    s953 & 1842.92 & 2531.02 & 38.61 & 1824.92 & 2372.20 & 29.99  & 23.76 & 36.65 & 54.25 & 23.76  & 33.41  & 40.59   \\ \hline
    s5378 & 676.40 & 729.87 & 7.9 & 676.40 & 729.87 & 7.9   & 10.17  & 10.36 & 1.82 & 10.17  & 10.36  & 1.82 \\ \hline
    s35932 & 38188.04 & 39631.27 & 3.78 & 38188.04 & 43058.6 & 12.75  & 767.28 & 791.8094 & 3.2 & 767.31  & 847.08  & 10.4    \\ \hline
    \textbf{Average} &  &  & 59.1 &  &  & 62.2 &  &  & 68.3 &  &  & 74.04  \\ \hline
    \end{tabularx}
    }
    \end{table*}
    
\begin{table}[htb!]
    \caption{Delay overhead analysis of SATConda with different encryption schemes for ISCAS'89 circuits}
    \label{tab:delay-overhead_89}
    \tcrevision{
    \begin{tabularx}{\columnwidth}{X|XXXXXX} 
    \hline
	&	\multicolumn{6}{c}{Delay Overhead } 	\\ \cline{1-7}
    Circuit & IOLTS- '14 \cite{IOLTS_14} $(ns)$ & SAT- Conda+\cite{IOLTS_14}  $(ns)$ & Over- head $(\%)$ & TOC'13- XOR \cite{toc13xor} $(ns)$ & SAT- Conda+\cite{toc13xor} $(ns)$ & Over- head $(\%)$  \\ \hline
    s382 & 1.2 & 1.93 & 60.83 & 1.2 & 1.67 & 39.17   \\ \hline
    s400 & 1.23 & 1.68 & 36.59 & 1.23 & 1.72 & 39.84  \\ \hline
    s526 & 1.59 & 1.66 & 4.4 & 1.59 & 1.95 & 22.64  \\ \hline
    s526n & 1.58 & 2.1 & 32.91 & 1.58 & 2.2 & 39.24  \\ \hline
    s641 & 3.3 & 3.58 & 8.48 & 3.3 & 3.7 & 12.12 \\ \hline
    s953 & 1.75 & 1.97 & 12.57 & 1.75 & 2.28 & 30.29  \\ \hline
    s35932 & 7.7 & 11.29 & 46.62 & 7.7 & 12.9 & 67.53  \\ \hline
    \textbf{Average} &  &  & 28.21 &  &  & 31.81     \\ \hline
    \end{tabularx}
    }
    \end{table}

\subsubsection{Overhead Analysis}
\rakib{
In addition to SAT-hardness, we evaluate the imposed overheads through the conversion. Table \ref{tab:overhead} reports the area and power overhead of the original encrypted circuit (\tcrevision{XOR/XNOR-based logic locking \cite{DAC_12}  and AND/OR-based logic locking \cite{IOLTS_14}}) and compare them with the overhead of the proposed SATConda (that successfully defends SAT-attacks, where the original \tcrevision{XOR/XNOR-based logic locking \cite{DAC_12}  and AND/OR-based logic locking \cite{IOLTS_14}} fail). }

\tcrevision{Both the DAC'12 logic locking  \cite{DAC_12}  and IOLTS'14 logic locking  \cite{IOLTS_14} introduce an average area overhead of 5\% when compared to the original circuit for the ISCAS'85 benchmarks. 
However, as seen earlier despite incurring such overheads, the SAT attack \cite{Subramanyan_HOST'15} can reverse engineer the 
IP. }
Further encrypting with SATConda incurs additional overhead of 
about 57\% and about 42\% on average compared to \cite{IOLTS_14} and \cite{DAC_12} respectively i.e., around 7.85\% and 7.1\% compared to original circuit, but guarantees security. 

In a similar manner, 
proposed SATConda incurs around 54\% and 36\% power overhead on an average compared to \cite{IOLTS_14} and \cite{DAC_12} respectively i.e., around 7.7\% and 6.8\% compared to original circuit.
As observed that the majority of the overhead comes from the base encryption technique (Say DAC'12 or IOLTS'14) rather than the proposed SATConda. 
It needs to be noted that the area overhead for other encryption techniques are reported in `Other Method' column, which 
when combined with SATConda can incur lower overheads compared to that with \cite{IOLTS_14}+SATConda. 
Thus, if a lightweight encryption which is moderate in terms of security exist, it can be made secure with SATConda with minimal overheads.

 
 
From Table \ref{tab:overhead} the area overhead follows another important trend. The relative area overhead due to our model follows a non-linear trend as the circuit size keeps increasing. For small circuits (i.e c432, c499 etc.) the area overhead is large compared to the area overhead for the large circuits such as c3540, c5315, and c7552. 
This depicts the scalability and better applicability for real-world larger circuits. 
For power overhead analysis, very similar trend is observed. 

\tcrevision{ Table \ref{tab:overhead_89} reports the area and power overhead of the ISCAS'89 encrypted circuit using (\tcrevision{ IOLTS'14 logic locking \cite{IOLTS_14} and TOC'13 logic locking \cite{toc13xor}}) with our proposed method. Here, the TOC'13 \cite{toc13xor}  and IOLTS'14 \cite{IOLTS_14} logic locking also introduce an average area overhead of 5\% when compared to the original circuit. Further encrypting with SATConda incurs additional area overhead of about 59\% and about 62\% on average compared to \cite{IOLTS_14} and \cite{toc13xor} respectively i.e., around 7.95\% and 8.1\% compared to original circuit. Similarly, proposed SATConda incurs around 68\% and 74\% power overhead on an average compared to \cite{IOLTS_14} and \cite{toc13xor} respectively i.e., around 8.4\% and 8.7\% compared to original circuit. Table \ref{tab:delay-overhead_89} reports the delay overhead of the ISCAS'89 encrypted circuit with our proposed method.
}

\rakibtc{We also evaluate the delay overhead for our proposed model. Table \ref{tab:delay-overhead} shows the delay overhead due to SATConda. With the increase in the original circuit size the delay overhead is also increased. However, our proposed model shows non-linear relation for delay overhead size as we keep increasing the circuit size. For example, the c3540 circuit has less delay overhead than the smaller c1908 circuit. We believe that the reason behind this lies in the input vs output ratio and the type of gate used for a given circuit.
Though c3540 circuit has almost double gate count than c1908 circuit, the number of output pin for c1908 is more than c3540. In order to deal with more output pin to make that circuit SAT-hard our model needs to add more clause which leads greater delay overhead. 
}

Based on the overall evaluations performed, we confirm that SATConda performs efficient translation of SAT to unSAT for both the ISCAS'85 and ISCAS'89 benchmarks with lower overhead, power consumption without deviating from the original functionality.


\begin{figure*}
  \centering
  \begin{subfigure}[b]{.45\linewidth}
    \includegraphics[width=\linewidth]{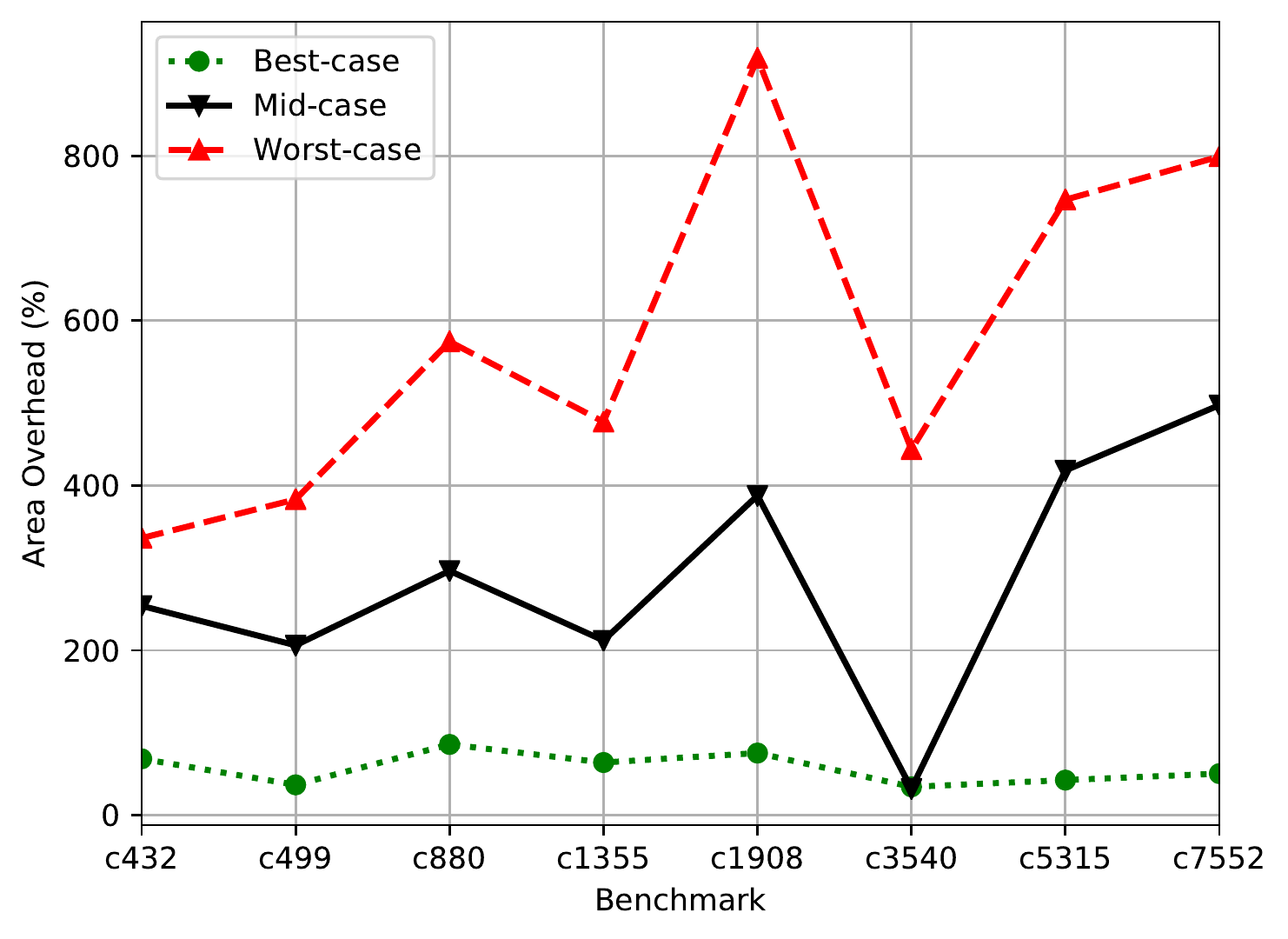}
    \caption{}
    \label{fig:area-overhead-iolts}
  \end{subfigure}
  \begin{subfigure}[b]{.45\linewidth}
    \includegraphics[width=\linewidth]{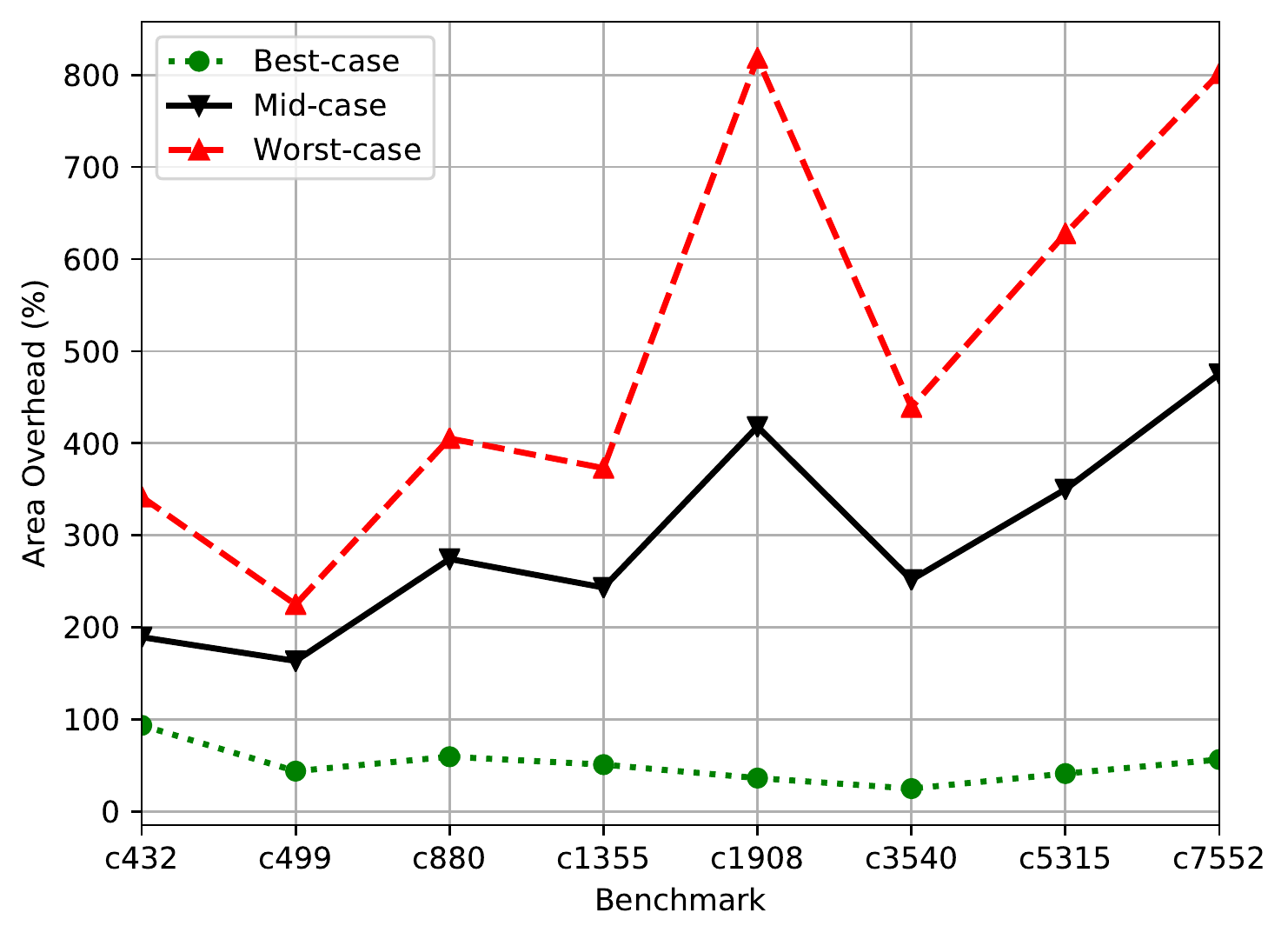}
    \caption{}
    \label{fig:area-overhead-dac}
  \end{subfigure}
  \caption{Area overhead analysis for ISCAS'85 benchmark circuit for different cases: (a) Area overhead on IOLTS'14 encryption \cite{IOLTS_14}, (b) Area overhead on DAC'12 encryption \cite{DAC_12}}
   \label{fig:area-overhead}
\end{figure*}

\begin{figure*}
  \centering
  \begin{subfigure}[b]{.45\linewidth}
    \includegraphics[width=\linewidth]{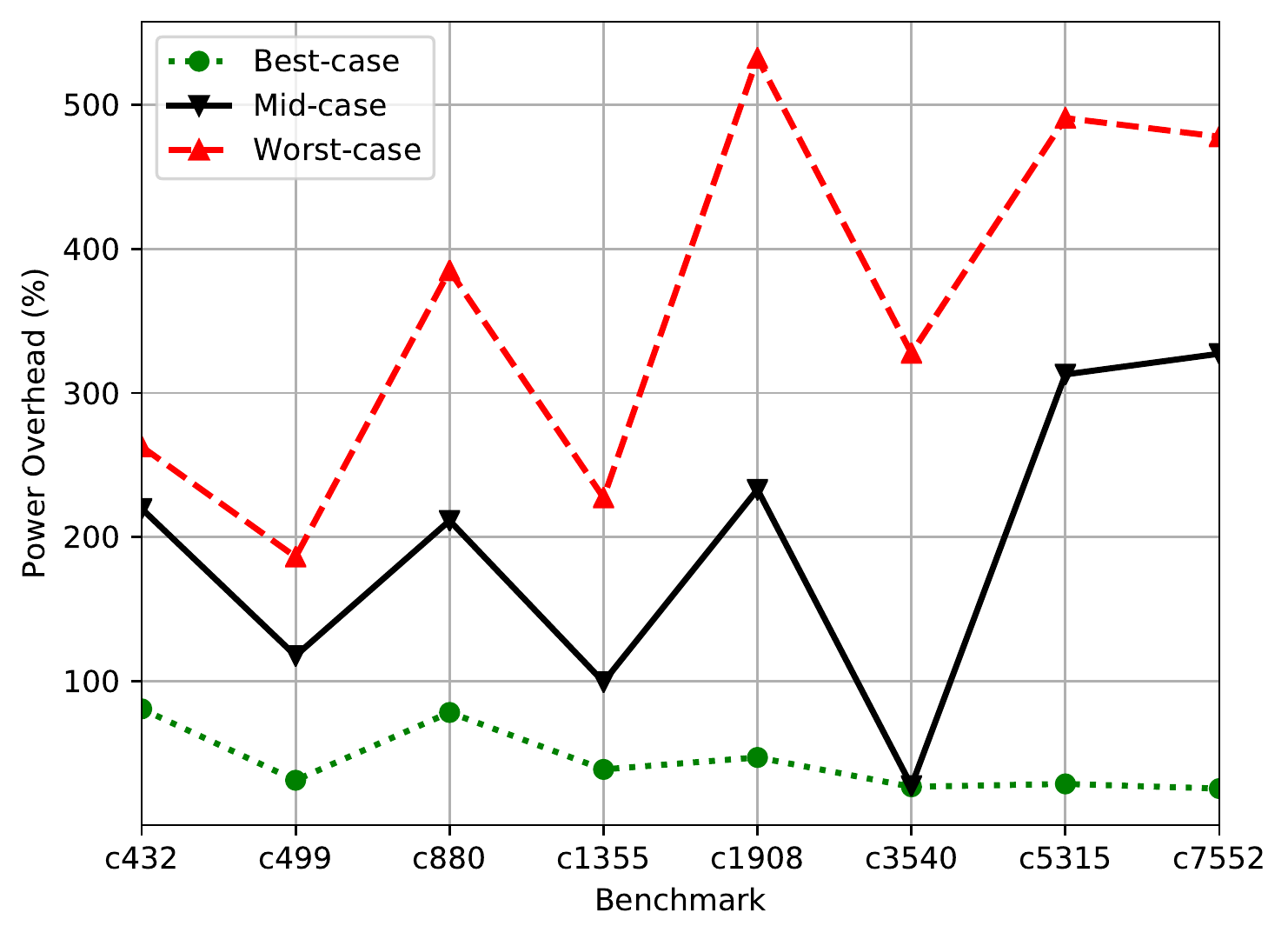}
    \caption{}
    \label{fig:power-overhead-iolts}
  \end{subfigure}
  \begin{subfigure}[b]{.45\linewidth}
    \includegraphics[width=\linewidth]{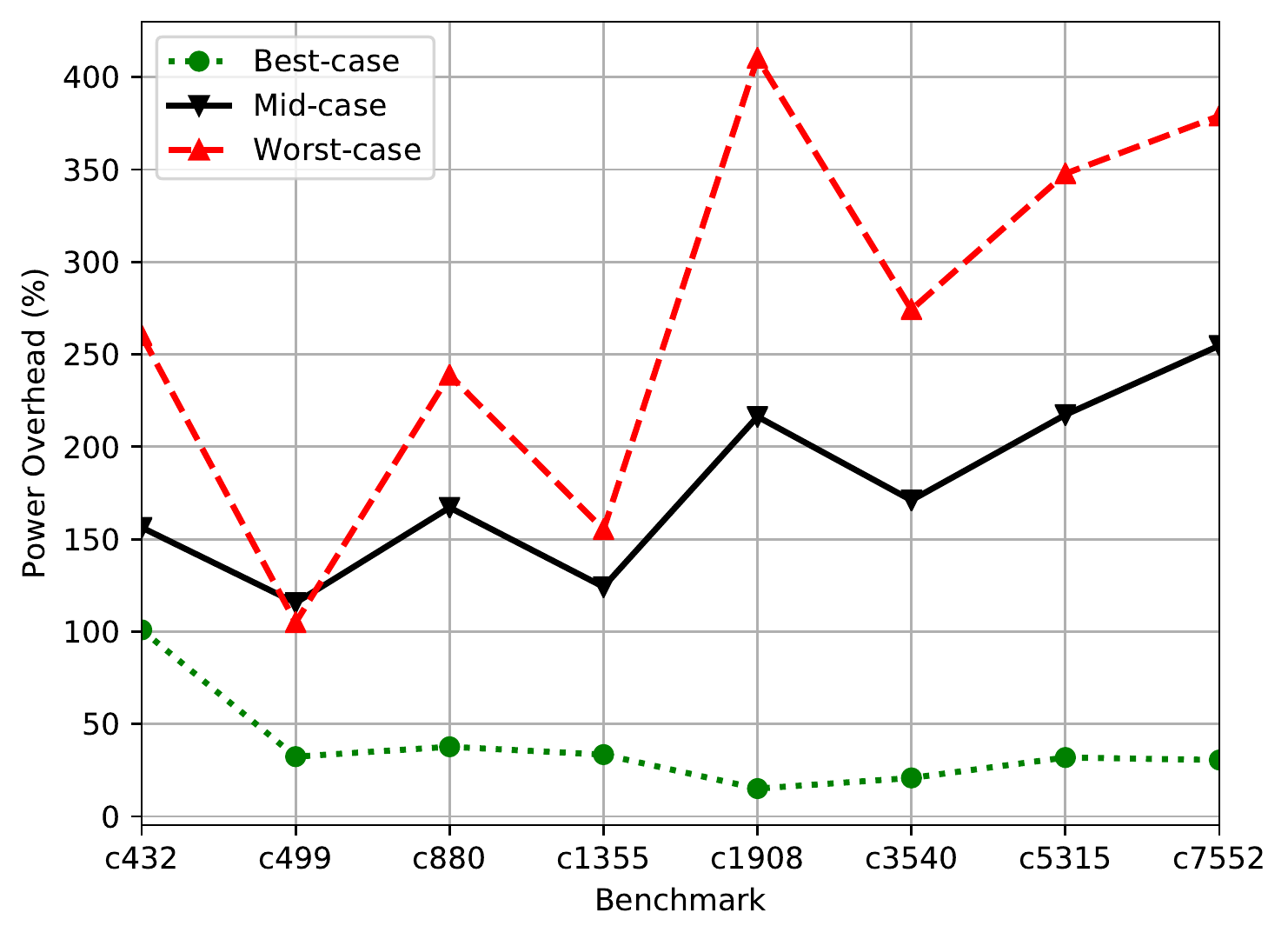}
    \caption{}
    \label{fig:power-overhead-dac}
  \end{subfigure}
  \caption{Power overhead analysis for ISCAS'85 benchmark circuit for different cases: (a) power overhead on IOLTS'14 encryption \cite{IOLTS_14}; (b) Power overhead on DAC'12 encryption \cite{DAC_12}}
   \label{fig:power-overhead}
\end{figure*}

\begin{figure*}
  \centering
  \begin{subfigure}[b]{.45\linewidth}
    \includegraphics[width=\linewidth]{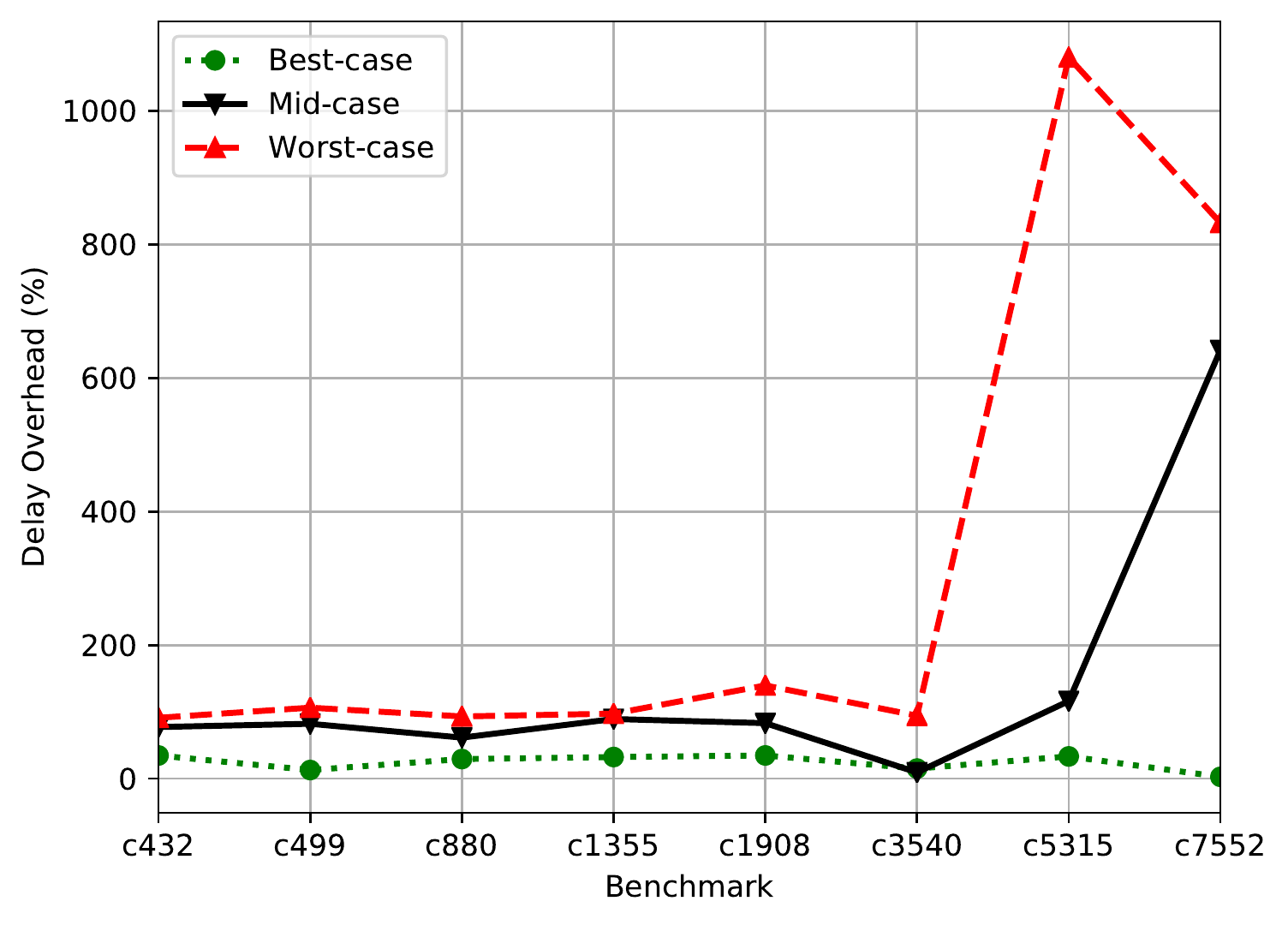}
    \caption{}
    \label{fig:delay-overhead-iolts}
  \end{subfigure}
  \begin{subfigure}[b]{.45\linewidth}
    \includegraphics[width=\linewidth]{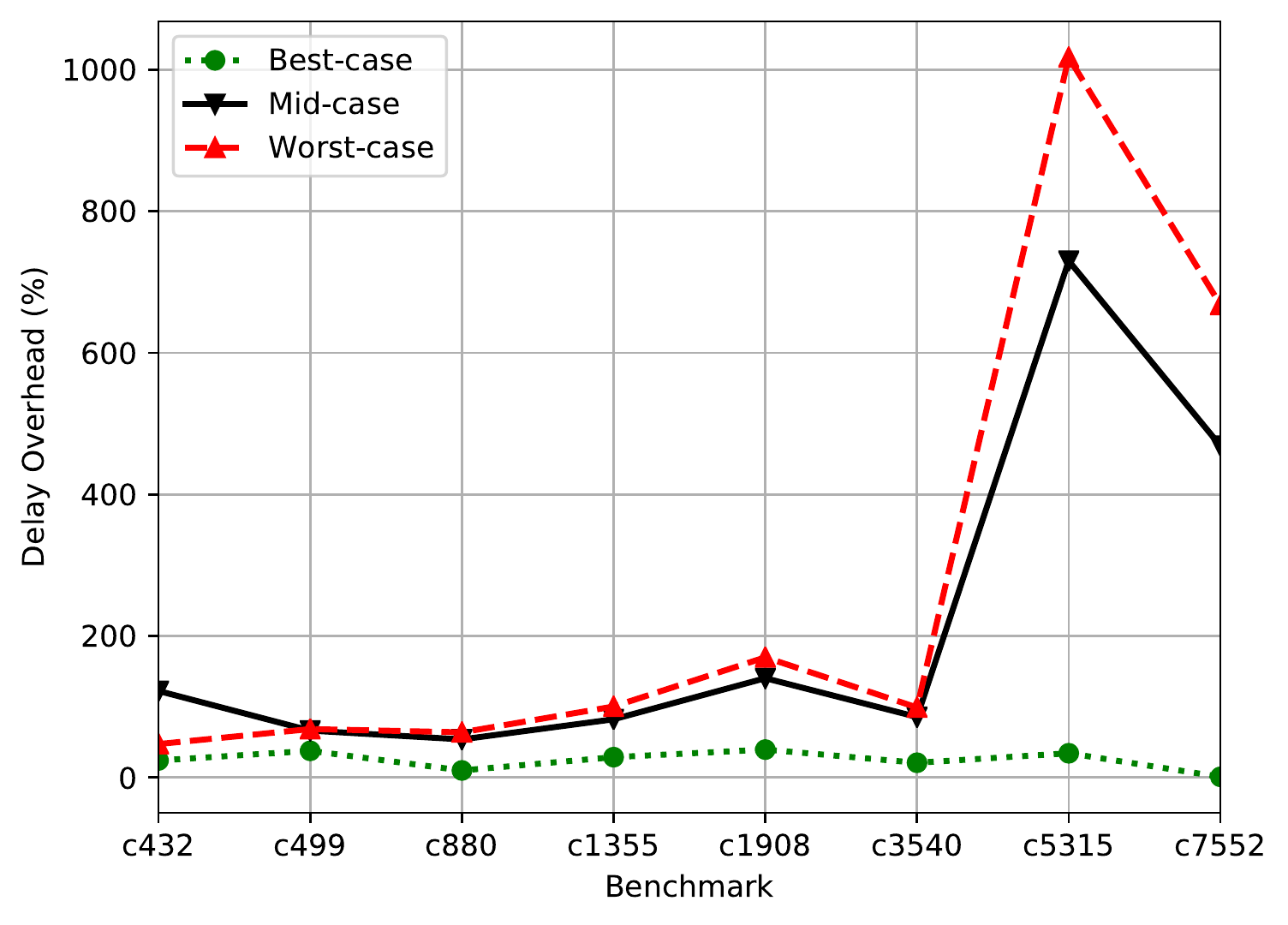}
    \caption{}
    \label{fig:delay-overhead-dac}
  \end{subfigure}
  \caption{Delay overhead analysis for ISCAS'85 benchmark circuit for different cases. (a) Delay overhead on IOLTS'14 encryption \cite{IOLTS_14}; (b) Delay overhead on DAC'12 encryption \cite{DAC_12}}
   \label{fig:delay-overhead}
\end{figure*}

\rakibtc{
\subsubsection{Impact of MPNN}
As the clause generation process is based on the model learnt by the MPNN, 
different training data can lead to different learnt models and different seed values. We analyze the impact of the seed on the 
overheads here, as all of them lead to unSAT in terms of security. 
We compare three different models that we achieved from SATConda. We named them as best-fit model (for $seed_1 = 0.9$ and $seed_2 = 0.9$ - leading to lowest overhead), mid-fit model(for $seed_1 = 0.5$ and $seed_2 = 0.5$), and worst-fit model(for $seed_1 = 0.3$ and $seed_2 = 0.4$).
Figure \ref{fig:area-overhead} depicts the area overhead (\%), Figure \ref{fig:power-overhead} shows the power overhead (\%), and Figure \ref{fig:delay-overhead} illustrates the delay overhead for different models. }

Figure \ref{fig:area-overhead} depicts the area overhead (\%) for ISCAS’85 benchmark circuit for different models where Figure \ref{fig:area-overhead-iolts} shows area overhead on IOLTS’14 encryption and Figure \ref{fig:area-overhead-dac} shows area overhead on DAC’12 encryption. 
As can be seen that the worst-fit model gives relatively large 
area overhead for all the benchmark circuits. The reason behind this is the worst-fit model adds a significant number of clauses to the original CNF file. Another important trend is observed from the Figure \ref{fig:area-overhead} that for best fit model the overhead is significantly lower and the overhead curve is almost horizontal than the mid-fit and the worst-fit model. On the other hand the worst-fit model shows significant fluctuations on different circuits as the model is not properly trained. 
On average, the area overhead for the best-fit, mid-fit, and the worst-fit model for the IOLTS'14 encryption is about $57$\%, $288$\%, and $584$\% respectively and for the DAC'12 encryption is about $51$\%, $295$\%, and $503$\% respectively , irrespective of the model fit used for generating or perturbting the clauses, each circuit shows security against SAT-attack.
 
Figure \ref{fig:power-overhead} depicts the power overhead (\%) for ISCAS’85 benchmark circuit for different models where Figure \ref{fig:power-overhead-iolts} shows power overhead on IOLTS’14 encryption and Figure \ref{fig:power-overhead-dac} shows power overhead on DAC’12 encryption. 
On average , the power overhead for the best-fit, mid-fit, and the worst-fit model for the IOLTS'14 encryption is about $44$\%, $190$\%, and $360$\% respectively and for the DAC'12 encryption is about $37$\%, $177$\%, and $271$\% respectively.
 
Figure \ref{fig:delay-overhead} depicts the power overhead (\%) for ISCAS’85 benchmark circuit for different models where Figure \ref{fig:delay-overhead-iolts} shows delay overhead on IOLTS’14 encryption and Figure \ref{fig:delay-overhead-dac} shows delay overhead on DAC’12 encryption. 
On an average, the delay overhead for the best-fit, mid-fit, and the worst-fit model for the IOLTS'14 encryption is about $24$\%, $145$\%, and $316$\% respectively and for the DAC'12 encryption is about $24$\%, $218$\%, and $275$\% respectively.
\tcrevision{A similar trend was observed for ISCAS'89 benchmark circuits on area, power, and delay overhead using our different models, not presented for the purpose of conciseness.}

From the above analysis, a straightforward observation can be made. The best-fit model shows lower area, power, and delay overhead compared to the other two cases. We have this best fit model when we train our MPNN with a large number of obfuscated circuit with large number of variables. In return, the best-fit model adds less number of clauses with minimum literals in each clause to meet the area, power, and delay overhead constraint.


\section{Conclusion}

In this work we successfully defend the popular SAT-attack by introducing a neural network based unSAT problem generator, SATConda, to achieve an enhanced obfuscation technique for hardware security regime. We observed that the integration of an unSAT block with a dummy output pin replacing an original output pin (keeping the same functionality) deceives the SAT-attack. Our framework is evaluated on the state-of-the-art benchmarks such as ISCAS'85 and ISCAS'89 using two state-of-the-art encryption algorithms. We validate our model with SAT-attack, AppSAT attack 
and three other traditional SAT solvers.


%





\ifCLASSOPTIONcaptionsoff
  \newpage
\fi



%




\bibliographystyle{IEEEtran} 
\bibliography{citation.bib}

%
\begin{IEEEbiography}[{\includegraphics[width=1in,height=1.25in,clip,keepaspectratio]{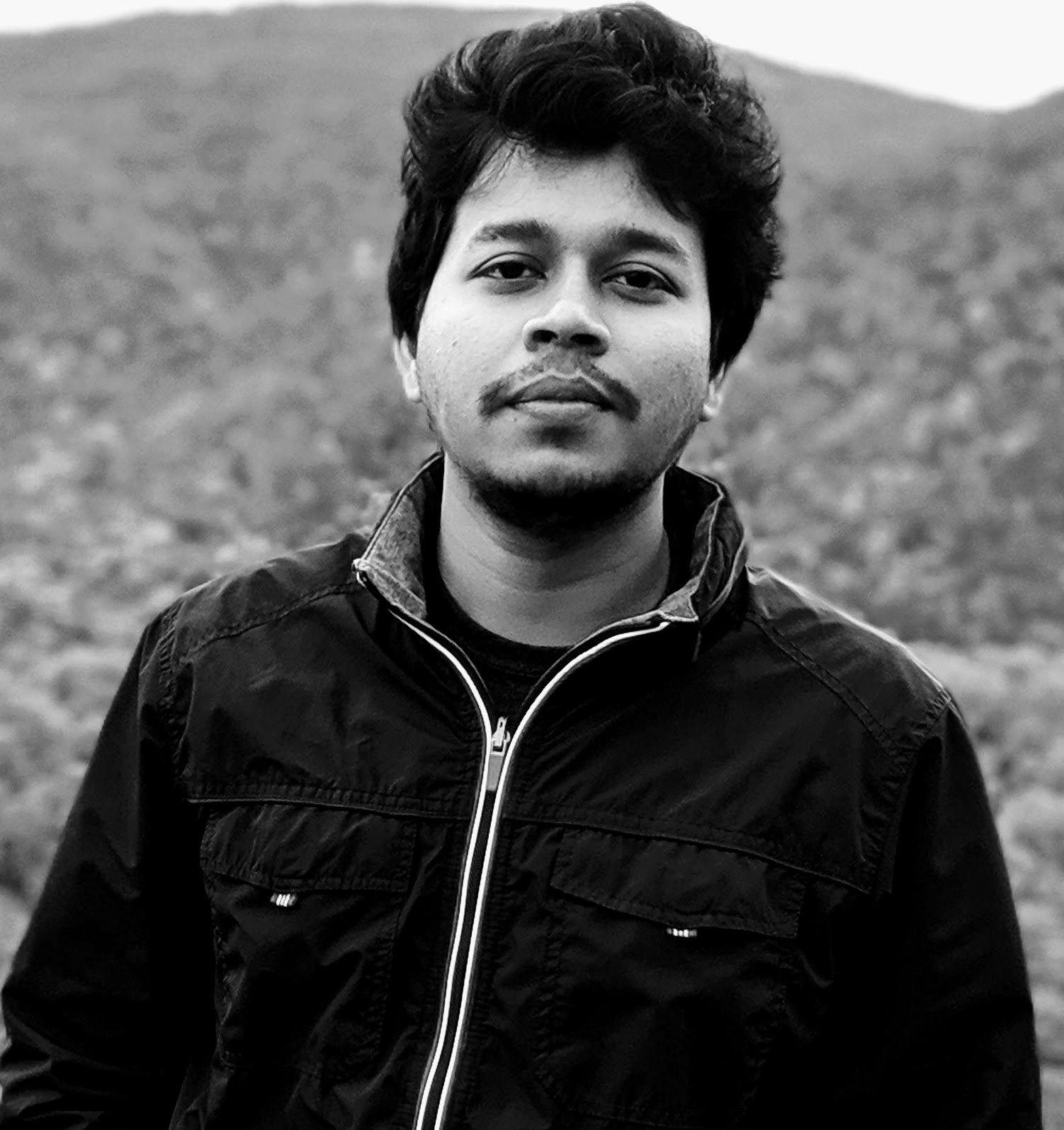}}]{Rakibul Hassan}
is a Ph.D student, currently conducting his research under the supervision of Dr. Sai Manoj P D, an Assistant professor at the Electrical and Computer Engineering Department, George Mason Universiy, Fairfax, VA, USA. Rakibul's present research interest includes computer architecture and IoT network security applying deep learning. He is also collaborating with another project, named, Adversarial attack on cloud computing. He has published his research work in the CASES’19 Companion conference and another one has been accepted in the ISQED’2020 conference. He  received his B.Sc. degree in electrical and electronic engineering from Ahsanullah University of Science and Technology, Dhaka, Bangladesh in 2016. After completing his bachelor degree he worked for two years as a lecturer at Bangladesh University, Dhaka, Bangladesh. During that time, he published several peer-reviewed conference papers and one journal paper in IEEE Transactions on Computer-Aided Design.
\end{IEEEbiography}

\begin{IEEEbiography}[{\includegraphics[width=1in,height=1.25in,clip,keepaspectratio]{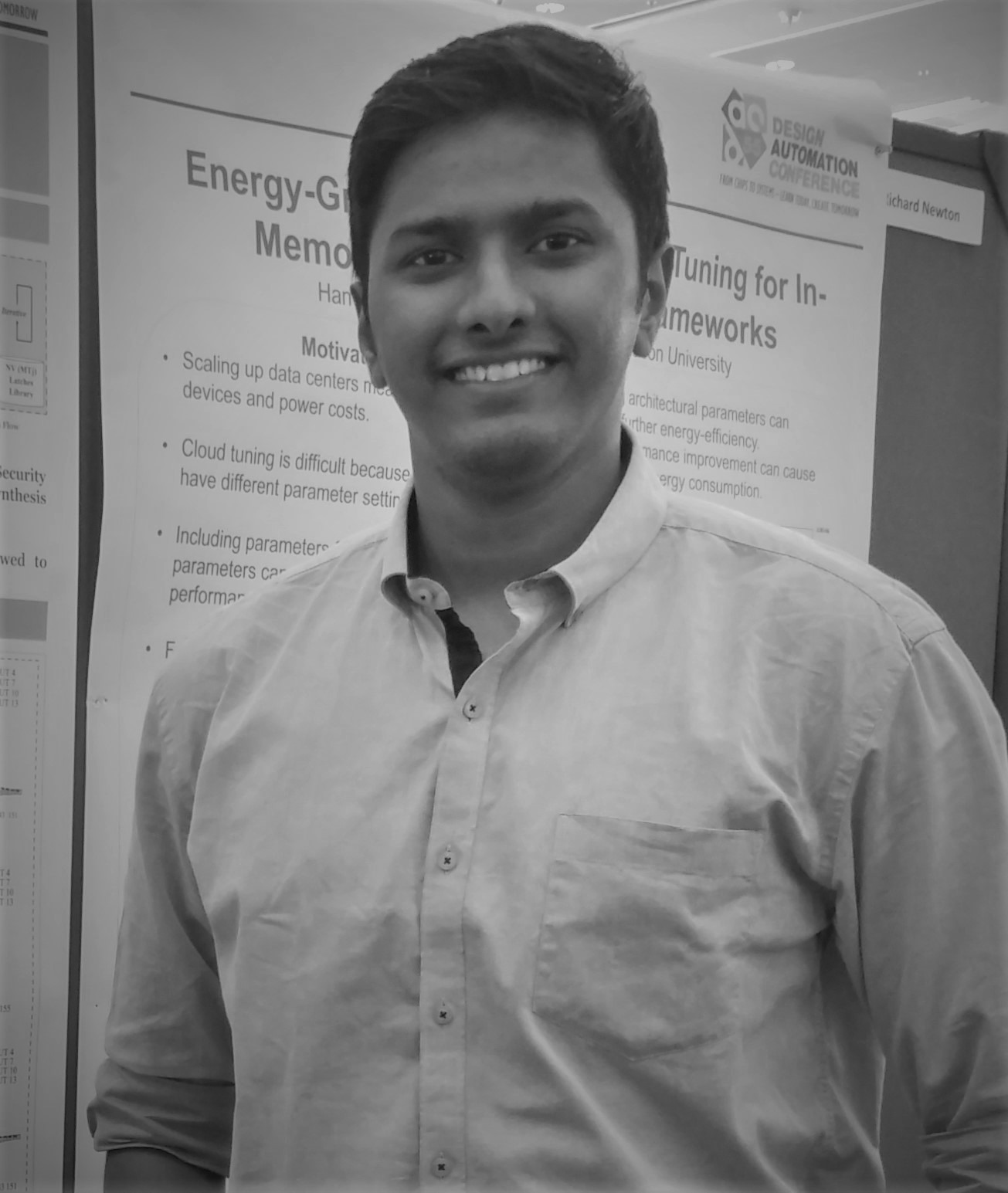}}]{Gaurav Kolhe}
is a Ph.D. student at Electrical and Computer Engineering Department of University of California, Davis,. His research interest are in the field of Heterogeneous Computing and Hardware Security and Trust, which spans the area of Computer Design and Embedded Systems. He is leading the research on "Hybrid Spin Transfer Torque-CMOS Technology to Prevent Design Reverse Engineering", a project funded by DARPA. He has previously worked for Information Sciences Institute, University of Southern California, where he had worked on DARPA's "Obfuscated Manufacturing for GPS (OMG)" program. He has received the "Richard Newton Fellowship Stud dent award 2018" at Design Automation Conference. Gaurav received his Master's degree in Computer Engineering in 2018 from George Mason University and BS degree in Electrical and Telecommunication Engineering in 2015 from Rajiv Gandhi College of Engineering, Nagpur, India.
\end{IEEEbiography}

\vspace{-2em}
\begin{IEEEbiography}[{\includegraphics[width=1in,height=1.25in,clip,keepaspectratio]{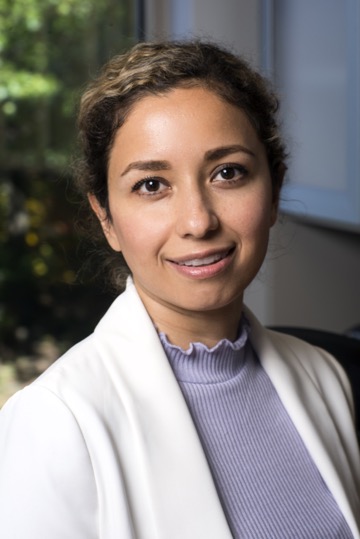}}]{Setareh Rafatirad}
is an Associate Professor in Department of
Information Sciences and Technology at George Mason University. She
obtained her M.Sc. and PhD in Computer Science from University of
California, Irvine in 2009 and 2012. She received the ICDM 2019 Best
Peper Award (9
Paper Award. She received research funding from government agencies
including NSF, DARPA, and AFRL for major projects. Her research
interest covers several areas including Big Data Analytics, Data
Mining, Knowledge Discovery and Knowledge Representation, IoT
Security,and Applied Machine Learning. Currently, she is actively
supervising multiple research projects focused on applying ML and Deep
Learning techniques on different domains including House Price
Prediction, Malware Detection, and Emerging big data application
benchmarking and characterization on heterogeneous architectures.
\end{IEEEbiography}
\vspace{-2em}
\begin{IEEEbiography}[{\includegraphics[width=1in,height=1.25in,clip,keepaspectratio]{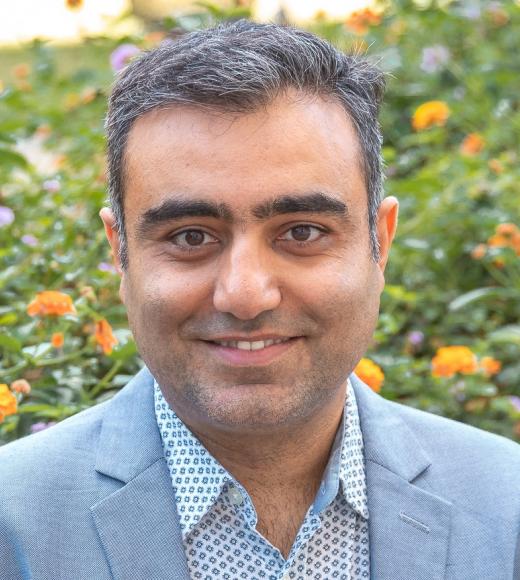}}]{Houman Homayoun}
is currently an Associate Professor in the Department of Electrical and Computer Engineering at University of California, Davis. Prior to that he was an Associate Professor in the Department of Electrical and Computer Engineering at George Mason University (GMU). From 2010 to 2012, he spent two years at the University of California, San Diego, as NSF Computing Innovation (CI) Fellow awarded by the CRA-CCC. Houman graduated in 2010 from University of California, Irvine with a Ph.D. in Computer Science. He was a recipient of the four-year University of California, Irvine Computer Science Department chair fellowship. Houman received the MS degree in computer engineering in 2005 from University of Victoria and BS degree in electrical engineering in 2003 from Sharif University of Technology. 
He is currently the director of UC Davis Accelerated, Secure, and Energy-Efficient Computing Laboratory (ASEEC).  Houman conduct research in hardware security and trust, data-intensive computing and heterogeneous computing, where he has published more than 100 technical papers in the prestigious conferences and journals on the subject and directed over \$8M in research funding from NSF, DARPA, AFRL, NIST and various industrial sponsors. He received several best paper awards and nominations in various conferences including GLSVLSI 2016, ICCAD 2019, and ICDM 2019. Houman served as Member of Advisory Committee, Cyber security Research and Technology Commercialization (R\&TC) working group in the Commonwealth of Virginia in 2018. Since 2017 he has been serving as an Associate Editor of IEEE Transactions on VLSI. He was the technical program co-chair of GLSVLSI 2018 and the general chair of 2019 conference.
\end{IEEEbiography}
\begin{IEEEbiography}[{\includegraphics[width=1.25in,height=1.25in,clip,keepaspectratio]{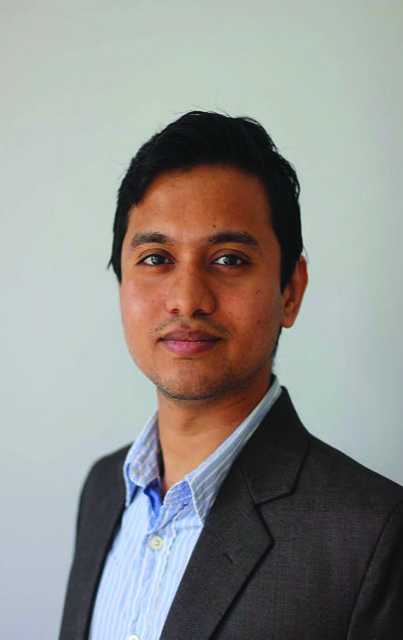}}]{Sai Manoj P D}
(S'13-M’15) is an assistant professor at George Mason University. Prior joining to George Mason University (GMU) as an assistant professor, he served as research assistant professor and post-doctoral research fellow at GMU and was a post-doctoral research scientist at the System-on-Chip group, Institute of Computer Technology, Vienna University of Technology (TU Wien), Austria. He received his Ph.D. in Electrical and Electronics Engineering from Nanyang Technological University, Singapore in 2015. He received his Masters in Information Technology from International Institute of Information Technology Bangalore (IIITB), Bangalore, India in 2012.
His research interests include on-chip hardware security, neuromorphic computing, adversarial machine learning, self-aware SoC design, image processing and time-series analysis, emerging memory devices and heterogeneous integration techniques. He won best paper award in Int. Conf. On Data Mining 2019, and his works were nominated for best paper award in prestigious conferences such as Design Automation \& Test in Europe (DATE) 2018, International Conference on Consumer Electronics 2020, and won Xilinx open hardware contest in 2017 (student category). He is the recipient of the ``A. Richard Newton Young Research Fellow'' award in Design Automation Conference, 2013.
\end{IEEEbiography}







\end{document}